\documentclass{article}
\usepackage{graphicx} 

\usepackage{fullpage} 

\usepackage{amsmath} 
\usepackage{amsthm}
\usepackage{amssymb}
\usepackage{stmaryrd}
\usepackage{xcolor}
\usepackage{tikz}
\usetikzlibrary{arrows.meta,calc}

\usepackage[inline]{enumitem}   
\makeatletter
\newcommand{\inlineitem}[1][]{%
\ifnum\enit@type=\tw@
{\descriptionlabel{#1}}
\hspace{\labelsep}%
\else
\ifnum\enit@type=\z@
\refstepcounter{@listctr}\fi
\quad\quad@itemlabel\hspace{\labelsep}%
\fi}
\makeatother

\newcommand{\out}{\mathrm{out}}
\newcommand{\inn}{\mathrm{in}}

\newcommand{\repr}{\operatorname{repr}}

\newcommand{\lrb}[1]{\llbracket#1\rrbracket}
\newcommand{\lbb}{\llbracket}
\newcommand{\rbb}{\rrbracket}
\newcommand{\N}{N}

\DeclareMathOperator{\dist}{dist}

\DeclareMathOperator{\SI}{SumIndex}
\DeclareMathOperator{\VSI}{VSI}

\newtheorem{theorem}{Theorem}[section]
\newtheorem{lemma}[theorem]{Lemma}
\newtheorem{corollary}[theorem]{Corollary}

\newtheorem{definition}[theorem]{Definition}
\newtheorem{observation}[theorem]{Observation}

\usepackage{ifpdf}
\ifpdf
    \usepackage{hyperref}
\else
    \usepackage[hypertex]{hyperref}
\fi

\usepackage[nameinlink,capitalise,noabbrev]{cleveref}

\title{Hardness Results for Approximate Distance Labeling\\ in Sparse Graphs\thanks{This work is partially supported by United States - Israel Binational Science Foundation (BSF) grant no. 2024233.
Contact: \texttt{kopelot.biu@gmail.com,  korinariel10@gmail.com, 
liamr@macs.biu.ac.il}.}}
\author{Tsvi Kopelowitz, Ariel Korin, Liam Roditty \\Bar-Ilan University, Ramat Gan, Israel}
\date{}

\begin{document}

\maketitle
\thispagestyle{empty}
\setcounter{page}{0}

\begin{abstract}
Distance labeling schemes assign short labels to the vertices of a graph so that distances between pairs of vertices can be recovered, exactly or approximately, from the two corresponding labels alone. For sparse graphs, Kosowski, Uzna\'nski, and Viennot [PODC'2019] showed that, assuming the hardness of the Sum-Index communication problem, exact distance labeling for sparse unweighted graphs requires near-linear label size. Their construction also gives lower bounds for stretch $1+o(1)$, but does not rule out substantially smaller labels for stretch $1+\varepsilon$, where $\varepsilon>0$ is fixed.

We prove lower bounds for this $(1+\varepsilon)$-stretch regime through a density/stretch tradeoff. For every constant integer $k\ge 1$, we show that any approximate distance labeling scheme with stretch strictly smaller than $1+1/k$ for unweighted graphs with $m=O(n^{1+1/k})$ edges requires maximum label size at least $\SI(n)/2^{O_k(\sqrt{\log n})}$ for infinitely many values of $n$. Thus, under the conjecture that $\SI(n)=n^{1-o(1)}$, the maximum label size must be $n^{1-o(1)}$. The main new ingredient is an unweighted graph construction based on $3$-AP-free sets, which replaces the weighted uniqueness mechanism of the previous Sum-Index reduction by an additive uniqueness mechanism and keeps the relevant critical paths short enough to obtain a constant multiplicative distance gap.
The same construction also yields near-tight lower bounds for related $2$-hop structures, including hub labeling schemes, shortcut sets, and approximate distance hopsets.

\end{abstract}


\section{Introduction}\label{sec:introduction}

A basic question in graph data structures is how much global information about a graph can be recovered from local information stored at its vertices.
This question is formalized by \emph{labeling schemes}, where each vertex is assigned a short binary label, and queries are answered using only the labels of the queried vertices.
Labeling schemes have been extensively studied, especially for adjacency, reachability, routing, and distance queries~\cite{moon1965minimal,kannan1992implicit,peleg2000proximity,graham1972embedding,gavoille2004distance,cohen2003reachability}.

Formally, let $U$ be a set whose elements represent the property of interest for a pair of vertices.
For example, $U$ may be $\{0,1\}$ for adjacency or reachability, or a set of numbers for distance estimation.
A labeling scheme  consists of a coder  and a decoder.
Given a graph $G=(V,E)$, the coder assigns to each vertex $v\in V$ a label $\ell(v)\in\{0,1\}^*$.
Given two labels $\ell(u)$ and $\ell(v)$, the decoder outputs an element of $U$.
The scheme is correct if, for every queried pair, the decoder outputs the corresponding value of the graph property.

\paragraph{Distance labeling schemes.}
In a distance labeling scheme, the decoder outputs an estimate $\hat d(u,v)\in \mathbb R_{\ge 0}\cup\{\infty\}$ for the graph distance $d_G(u,v)$.
For a parameter $t\ge 1$, we say that the scheme has stretch $t$ if, for every pair of vertices $u,v$,
$$
d_G(u,v) \le \hat d(u,v) \le t\cdot d_G(u,v).
$$
The case $t=1$ is exact distance labeling, while the case $t>1$ is approximate distance labeling.

The space of a labeling scheme is measured over a graph family $\mathcal G$, assuming that the scheme satisfies the required guarantee for every $G\in\mathcal G$.
The two standard measures are the maximum label size, $\max_{G\in\mathcal G}\max_{v\in V(G)}|\ell_G(v)|$, and the total space, $\max_{G\in\mathcal G}\sum_{v\in V(G)}|\ell_G(v)|$.

\paragraph{Exact distance schemes in general graphs.}
For general unweighted undirected graphs, exact distance labeling is well understood up to constant factors.
A classical counting lower bound gives an $\frac{n}{2}-O(1)$ bit lower bound on the maximum label size~\cite{gavoille2004distance}.
The current best upper bound is $\frac{\log_2 3}{2}n+o(n)$ bits, with constant decoding time~\cite{alstrup2016simpler}, improving earlier $O(n)$-bit schemes~\cite{graham1972embedding,gavoille2004distance,weimann2011note}.

Thus, for general graphs, exact distance labels must have linear size.
Two natural ways to seek smaller labels are to allow approximation, or to restrict attention to sparse graphs.

\paragraph{Approximate distance schemes in general graphs.}
For approximate distances, a fundamental result of Thorup and Zwick~\cite{thorup2005approximate} states that for every integer $k\ge 1$ there is a labeling scheme for undirected graphs with stretch $2k-1$ and total size $\tilde O(n^{1+1/k})$\footnote{$\tilde O$ suppresses polylogarithmic factors.}.
Furthermore, this bound is essentially tight, even for \emph{centralized} data structures, under the \emph{Erd\H{o}s girth conjecture}~\cite{thorup2005approximate}.
Additional approximate labeling schemes were developed for special graph classes and for hybrid notions of approximation~\cite{gavoille2001approximate,abraham2011approximate}.

For directed graphs, any finite-stretch distance labeling scheme is also a reachability labeling scheme, which requires $n/4-o(n)$ bits per label~\cite{bonamy2021optimal}.
Thus, the interesting setting for sublinear-size approximate distance labels is undirected graphs.

\paragraph{Schemes for sparse graphs.}
The sparsity relaxation is less understood.
Consider, for example, the family of unweighted undirected graphs satisfying $m=O(n)$.
On the upper bound side, there are exact distance labeling schemes that assign $o(n)$ bits per label for such graphs~\cite{alstrup2016sublinear,gawrychowski2016sublinear,kosowski2019hardness}, but no bound of the form $O(n^{1-\varepsilon})$ is known to hold for any constant $\varepsilon>0$.
On the lower bound side, the best unconditional lower bound on the maximum label size is $\Omega(\sqrt n)$~\cite{gavoille2004distance}.
In fact, the same $\Omega(\sqrt n)$ lower bound is known even in the case where a fixed additive error is allowed~\cite{gavoille2001approximate, alstrup2016simpler}.

\paragraph{Sum-Index.}
In an attempt to close this gap, Kosowski, Uzna\'nski, and Viennot~\cite{kosowski2019hardness} related distance labeling on sparse graphs to the communication complexity of Sum-Index.

\begin{definition}[Sum-Index]
The Sum-Index problem is a $3$-party communication problem, where Alice and Bob are both given the same binary string $S \in \{0,1\}^n$, for $n\in \mathbb{N}$. In addition, Alice is given an index $i \in \lrb n$\footnote{We use $\lrb n$ to denote $\{0,\dots,n-1\}$.}, and Bob is given an index $j \in \lrb n$. Alice and Bob each send one message to a referee, who must output $S[(i+j)\bmod n]$.
For a protocol $\Pi$, let $M_A(\Pi)$ and $M_B(\Pi)$ denote the maximum message lengths, in bits, sent by Alice and Bob, respectively, over all inputs. The communication complexity of Sum-Index is
$$
\SI(n)=\min_{\Pi} \max\{M_A(\Pi),M_B(\Pi)\},
$$
where the minimum is over all correct protocols for the problem.
\end{definition}

Similarly to exact distance labeling schemes in sparse graphs, Sum-Index has a simple $\Omega(\sqrt n)$ lower bound~\cite{nisan1993rounds,babai1995simultaneous,pudlak1997boolean,babai2003communication}, but no $O(n^{1-\varepsilon})$ upper bound is known. The best known upper bound is $\SI(n)=O(\frac{n \log^{1/4}n}{2^{\sqrt{\log n}}})$~\cite{ambainis1996upper}.

Kosowski, Uzna\'nski, and Viennot~\cite{kosowski2019hardness} proved that, already for unweighted undirected sparse graphs of maximum degree $3$, distance labeling schemes require maximum label size $\Omega\left(\frac{\SI(n)}{2^{\Theta(\sqrt{\log n})}}\right)$ for exact distances, and also for stretch $1+o(1)$.
Their result gives a concrete barrier for any major improvement in exact distance labeling schemes for sparse graphs.

However, their lower bound does not apply to stretch $1+\varepsilon$ for any fixed constant $\varepsilon>0$.
The reason is that the construction underlying their reduction is weighted, and their unweighted lower bound is obtained by replacing weighted edges with paths.
This preserves the exact-distance structure, but it makes the relevant distances large compared to the additive gap used in the reduction.
Consequently, the resulting multiplicative gap tends to $1$.

\subsection{Our Results}

The remaining question is whether Sum-Index hardness can give near-linear lower bounds for sparse unweighted graphs when the stretch is bounded away from $1$.
We answer this question for a density/stretch tradeoff.
For every constant integer $k\ge 1$, we prove a lower bound for unweighted graphs with $m=O(n^{1+1/k})$ edges and stretch strictly smaller than $1+1/k$.

\begin{theorem}\label{thm:distance-labeling-final-bound}
For every constant integer $k\ge 1$, there exists a constant $C_k>0$ such that for infinitely many integers $n\ge 1$, every approximate distance labeling scheme with stretch strictly smaller than $1+1/k$ for unweighted $n$-vertex graphs with $m=O(n^{1+1/k})$ edges requires maximum label size at least
$$
\Omega\left(\frac{\SI(n)}{2^{C_k\sqrt{\log n}}}\right).
$$
\end{theorem}

Thus, under the conjecture that $\SI(n)=n^{1-o(1)}$, \cref{thm:distance-labeling-final-bound} gives an $n^{1-o(1)}$ lower bound on the maximum label size for these graph families.
The theorem should be viewed as a tradeoff between density and stretch. As $k$ grows, the edge bound $m=O(n^{1+1/k})$ becomes closer to linear, while the stretch threshold $1+1/k$ approaches $1$.
For every fixed $k$, however, the stretch threshold is a fixed constant larger than $1$, so this goes beyond the $1+o(1)$ regime obtained from the unweighted conversion of~\cite{kosowski2019hardness}.

Our main new ingredient is the graph construction.
The construction of Kosowski, Uzna\'nski, and Viennot~\cite{kosowski2019hardness} uses weights to create the unique shortest paths needed by the Sum-Index reduction.
We replace this weighted uniqueness mechanism by an unweighted additive one, based on subsets of $\{1,\dots,N\}$ containing no nontrivial $3$-term arithmetic progression.
Throughout this paper we call such sets \emph{$3$-AP-free sets}.
This gives short unweighted critical paths of length exactly $2k$, and hence a distance gap between $2k$ and at least $2k+2$.

The quantitative form of our bound is governed by $r_3(N)$ which is the maximum size of a $3$-AP-free set taken from $\{1,\ldots N\}$.
The following more detailed theorem is our main technical lower bound.

\begin{theorem}\label{thm:distance-labeling-sumindex}
For every integer $k\ge 1$ and all sufficiently large $N$, every approximate distance labeling scheme for unweighted undirected graphs with $n=(2k+1)(9N)^k$ vertices and $m=O(n^{1+1/k})$ edges, and stretch strictly smaller than $1+1/k$, requires maximum label size
$$
\Omega\left(
\frac{r_3(N)^k}{2^kN^k\log(N^k)}
\cdot
\SI(N^k)
\right).
$$
\end{theorem}

Using the known lower bounds on $r_3(N)$ stated in \cref{thm:r3-known-bounds}, \cref{thm:distance-labeling-sumindex} implies \cref{thm:distance-labeling-final-bound} (see Section~\ref{sec:proof_main_lb_thm}).
In this sense, after substituting the current bounds for $r_3(N)$, we obtain the same type of Sum-Index loss as in~\cite{kosowski2019hardness}, namely a factor of the form $2^{O(\sqrt{\log n})}$.
Thus, any improvement in the known lower bounds for $r_3(N)$ would translate directly into an improved labeling lower bound.

\paragraph{Consequences for 2-hop and hub labeling.}
The new graph construction we use in Section~\ref{sec:construction} has consequences for  \emph{hub-labeling} (also known as $2$-hop labeling)~\cite{abraham2011approximate, kosowski2019hardness}, for shortcut sets and for approximate distance hopsets. 
These consequences are stated and proved in \cref{sec:consequences}.
We refer to Section~\ref{sec:extraPreliminaries} for the relevant definitions.
In the description below, we explain our results in the terminology of shortcut sets. Specifically, our results imply lower bounds on the size of shortcut sets
that reduce the reachability diameter (henceforth, diameter) of a graph
from $2k$ to at most $2$. However, we prove the same bounds for
approximate distance hopsets with stretch strictly smaller than $1+1/k$ that reduce the hopbound to $2$ and for the corresponding hub-labeling schemes (see Section~\ref{sec:immediate_consequences}). 

We start by providing some background, to place our results in the broader context of shortcut sets research.
In a breakthrough result, Hesse~\cite{hesse2003directed} showed that for any constant $\varepsilon>0$ there exists a constant $\delta=\delta(\varepsilon)>0$ and graphs $G$ with $O(n^{1+\varepsilon})$ edges, such that for any shortcut set $H$ of size $O(n^{2-\varepsilon})$, the diameter of the augmented graph $G\cup H$ remains $\Omega(n^\delta)$.
In recent years, much effort has been devoted to understanding the exact 
strength of shortcut sets~\cite{HuangP18, kogan2022new,
bodwin2023folklore,
williams2024simpler,
hoppenworth2025new}. For example, Vassilevska Williams, Xu and Xu~\cite{williams2024simpler} showed that the result stated above by Hesse~\cite{hesse2003directed} also holds for sparse graphs with $m=O(n)$.
Since such graphs $G$ must have an original diameter of $\Omega(n^\delta)$, a question one might ask is what happens when we restrict the initial diameter of such graphs. 

Specifically, consider the case where we are given a graph $G$ with bounded diameter $2k$, for a constant integer $k\ge 2$, and we are interested in a shortcut set $H$ which will reduce the diameter of $G\cup H$ to $2$.
Berman, Raskhodnikova and Ruan~\cite{BermanRR10} considered a closely related problem\footnote{Berman, Raskhodnikova and Ruan~\cite{BermanRR10} were interested in \emph{transitive closure spanners}, which are strongly related to shortcut sets, so their result uses a different terminology to the one stated here.}, and showed a construction, inspired by the one of Hesse~\cite{hesse2003directed}, which implies a lower bound for such graphs $G$ with $m=O(n^{1+1/k})$ of $|H|=\Omega(n^{2-\varepsilon})$ for any constant $\varepsilon>0$.
In this paper, we study the
subpolynomial loss from $n^2$ of this lower bound, and study the more general sparsity regime $m=O(n^{1+\alpha})$,
where $0\le\alpha\le1/k$.

For the setting implied by the work of Berman, Raskhodnikova and Ruan~\cite{BermanRR10}, we show in \cref{cor:2-shortcutSetLB} that for any constant integer $k\ge 2$, there are graphs $G$ with $m=O(n^{1+1/k})$ edges and
diameter $2k$, such that every shortcut set $H$ for which $G\cup H$ has
diameter at most $2$ must have size at least
\[
\Omega\!\left(n\cdot r_3(n^{1/k})^k\right)
\ge
\frac{n^2}{2^{\Theta_k(\sqrt{\log n})}}.
\]
We also show a near-tight lower bound for the density $m=O(n)$ (a simple upper bound is provided in Section~\ref{sec:upperBounds}. See discussion in Section~\ref{sec:discussion}).
Specifically, applying the directed version of the construction
underlying \cref{cor:approx-hubLabeling-tradeoff} with base
parameter $k-1$ and $\alpha=0$, we obtain, for every fixed integer
$k\ge3$, directed graphs with $m=O(n)$ edges and diameter $2k$, such that every shortcut set $H$ for which $G\cup H$ has
diameter at most $2$ must have size at least
\[
\Omega\!\left(
n\cdot r_3\!\left(n^{1/k}\right)^{k-1}
\right)
=
n^{2-1/k-o(1)}.
\]

Similar lower bounds are obtained for approximate distance hopsets in \cref{cor:approx-hubLabeling-tradeoff} and for hub labeling in \cref{cor:unweighted-reachability-hubLabeling}.

\subsection{Technical overview}

\paragraph{The proof template.}
We follow the communication-complexity strategy of Kosowski, Uzna\'nski, and Viennot~\cite{kosowski2019hardness}. A distance labeling scheme with small labels is used to obtain a low-communication protocol for Sum-Index. The high-level reduction template is therefore not new. The main question is what kind of graph construction can make this template work for unweighted graphs and for stretch bounded away from $1$. The formal lower bound obtained from this approach is stated in \cref{thm:distance-labeling-sumindex}, and \cref{thm:distance-labeling-final-bound} is then derived from it.

To turn such a graph into an approximate-distance lower bound, the desired distance gap is simple. For many pairs of endpoints, the distance should be $2k$ in one case, and at least $2k+2$ in the other. A distance decoder with stretch strictly smaller than $1+1/k$ can distinguish these two distances, since an estimate for a true distance $2k$ must be smaller than $(1+1/k)2k=2k+2$, whereas an estimate for a true distance at least $2k+2$ must be at least $2k+2$.

Thus, the graph-theoretic approach is to start with a base graph $G_0$ that contains many short unweighted unique shortest paths of length $2k$, which we call \emph{critical paths}. Intuitively, the endpoints of these paths represent possible inputs of Alice and Bob. We denote the possible Alice endpoints by $\mathcal X_{\mathrm A}$ and the possible Bob endpoints by $\mathcal X_{\mathrm B}$. In the idealized picture, if $x\in\mathcal X_{\mathrm A}$ represents Alice's index $i$ and $y\in\mathcal X_{\mathrm B}$ represents Bob's index $j$, then the critical path associated with this endpoint pair tests the position $(i+j)\bmod |S|$ of the common string.

Given the common string $S$, we define a subgraph $G_S$ of $G_0$ so that, intuitively, every critical path that is used by the reduction tests one bit of $S$. The path remains in $G_S$ if and only if the bit it tests is $1$. Hence the distance between the endpoints is $2k$ in the $1$ case, and at least $2k+2$ in the $0$ case.

The construction in~\cite{kosowski2019hardness}, reviewed in \cref{sec:KUV}, creates the critical paths using weights on the edges. This weighted construction gives the uniqueness needed for the reduction.

The same work also converts this construction to an unweighted graph by replacing weighted edges with paths. This is sufficient for their exact-distance lower bound, but for our approximate-distance goal the resulting distances are too large relative to the additive gap, and the stretch lower bound obtained in this way is only of the form $1+o(1)$. Our contribution is to replace the weighted uniqueness mechanism by an unweighted one, while preserving enough of the Sum-Index structure to obtain a strong lower bound.

\paragraph{Implementing the encoding.}
We now make the preceding description more precise. The graph $G_0$ is layered with $2k+1$ layers. The possible Alice endpoints $\mathcal X_{\mathrm A}$ lie in the first layer, and the possible Bob endpoints $\mathcal X_{\mathrm B}$ lie in the last layer. We call an endpoint pair $(x,y)\in\mathcal X_{\mathrm A}\times\mathcal X_{\mathrm B}$ \emph{supported} if the construction designates a critical path from $x$ to $y$.

Every  critical path has a distinguished vertex in the middle layer. This vertex is the place where the path records which position of the common string it tests. Thus, if the critical path from $x$ to $y$ passes through the middle vertex representing position $q$, we say that the supported endpoint pair $(x,y)$ tests $q$.

For each such position $q$, let $b_S(q)$ denote the bit of $S$ associated with $q$. The subgraph $G_S$ is obtained from $G_0$ by keeping the relevant middle-layer edges for positions $q$ with $b_S(q)=1$, and deleting them\footnote{
In~\cite{kosowski2019hardness}, the corresponding
middle-layer vertices are deleted. We instead delete all
their incident edges while keeping the vertices.
For the distance tests used in the reduction this has the same effect.
The reason for this change is aesthetic: our modification ensures that every graph in the family has exactly the same number $n$ of vertices, and we find the theorem more clearly stated this way.
} for positions $q$ with $b_S(q)=0$. Therefore, if a supported endpoint pair $(x,y)$ tests $q$, then the critical path from $x$ to $y$ remains in $G_S$ if and only if $b_S(q)=1$. Since this path is the unique path of length $2k$ between $x$ and $y$ in $G_0$, deleting its middle part destroys all paths from $x$ to $y$ of length $2k$. This gives exactly the distance gap described above. The formal implementation of this encoding appears in the proof of \cref{thm:distance-labeling-sumindex}.

\paragraph{The induced protocol.}
The communication protocol now follows directly. Since $S$ is common, both Alice and Bob know the graph $G_S$. Given their inputs, Alice chooses an endpoint $x\in\mathcal X_{\mathrm A}$ and Bob chooses an endpoint $y\in\mathcal X_{\mathrm B}$ so that the supported pair $(x,y)$ tests the position whose bit must be returned. Alice sends the label of $x$ in $G_S$, and Bob sends the label of $y$ in $G_S$. The referee applies the distance decoder. By the distance gap above, the decoder output determines whether the tested bit is $1$ or $0$. See \cref{sec:generalLBs}.

\paragraph{How the previous construction fits the template.}
The construction in~\cite{kosowski2019hardness} fits this template very naturally. In the weighted version of their construction, the relevant endpoint pairs are highly structured. Informally, Alice's input is embedded as an even coordinate $2i$, Bob's input is embedded as an even coordinate $2j$, and the critical path is arranged so that its middle vertex has coordinate $(2i+2j)/2=i+j$. Thus, the middle vertex directly represents the Sum-Index address. This is why their reduction can work with ordinary Sum-Index, up to a padding loss. We recall this construction in \cref{sec:KUV}.

The role of weights in their construction is to force the desired shortest paths to be unique. Our task is to obtain the same kind of behavior without weights, while keeping the critical paths short enough to give the constant stretch gap described above.

\paragraph{The unweighted uniqueness mechanism.}
Our construction replaces the weighted uniqueness mechanism by an additive uniqueness mechanism. We use a $3$-AP-free set $A$, meaning a set with no nontrivial three-term arithmetic progression. Equivalently, if $x,s,y\in A$ and $x+y=2s$, then $x=y=s$. The required background on $3$-AP-free sets and the known bounds on $r_3(N)$ are given in \cref{sec:preliminaries}, especially \cref{thm:r3-known-bounds}.

The graph is layered, and along a critical path each coordinate is incremented twice. Suppose that in one coordinate the intended increment is $s\in A$, so that the total increment in this coordinate is $2s$. Any other path of length $2k$ with the same endpoints would have to use two increments $x,y\in A$ satisfying $x+y=2s$. Since $A$ is $3$-AP-free, this forces $x=y=s$. Applying this argument in every coordinate shows that the path is the unique shortest path between its endpoints.

Thus, the graph is unweighted, but still has many unique shortest paths of length exactly $2k$. This is the replacement for the weighted uniqueness mechanism in~\cite{kosowski2019hardness}. The resulting layered construction is stated formally in \cref{thm:avgfree-layered-CP} in \cref{sec:construction}.

\paragraph{The structure compared to~\cite{kosowski2019hardness}.}
The price is that our supported endpoint pairs are less dense. In~\cite{kosowski2019hardness}, after choosing the appropriate endpoint sets, the construction supports a product-like family of pairs. Alice chooses her endpoint, Bob chooses his endpoint, and the graph supplies the corresponding critical path. In our construction, a pair of endpoints is supported only when the coordinatewise difference of the input vectors of the players lies in $D=A^k$.
Thus, as opposed to the construction in~\cite{kosowski2019hardness}, our construction does not seem to work with the original Sum-Index problem. 

To address this gap, in our lower bound proof we use a variant of Sum-Index that captures the nature of the supported pairs in our construction.
In~\cref{sec:sumindexreduc} we reduce the original Sum-Index problem to our new variant, which causes the main efficiency loss. If the ambient coordinate box has side length $N$, then allowing only differences from $A^k$ keeps roughly a $(|A|/N)^k$ fraction of the possible differences. Equivalently, reducing back to ordinary Sum-Index costs roughly a factor $(N/|A|)^k$. The reductions also introduce an additional factor $2^k$ and a logarithmic loss. This is why the final lower bound in \cref{thm:distance-labeling-sumindex} contains the factor $|A|^k/(2^kN^k\log(N^k))$, and after choosing $A$ to be a largest $3$-AP-free set, the factor $r_3(N)^k/(2^kN^k\log(N^k))$.

\section{Preliminaries}\label{sec:preliminaries}

\paragraph{Graphs.}
Let $G=(V,E)$ be a possibly weighted, possibly directed graph. We write
$V(G)=V$ and $E(G)=E$. 
For vertices $u,v\in V$, we write $\dist_G(u,v)$ for the length
of the shortest path from $u$ to $v$ in $G$, and set $\dist_G(u,v)=\infty$ if
no such path exists. 
We omit the subscript $G$ when the graph is clear
from context.
For a parameter $k$, we use asymptotic notations such as $O_k(\cdot)$ and $\Omega_k(\cdot)$ to hide factors depending on $k$. When $k$ is constant, we occasionally omit the subscript $k$.

\paragraph{Vectors.}

For an integer $N\ge 1$ and vector $\vec{x}=(x_0,\dots,x_{k-1})\in \mathbb{Z}^k$, let
$
\repr_N(\vec{x}) := \sum_{b=0}^{k-1} x_b N^b
$
be the number which $\vec x$ represents in base $N$, where the significance of the coordinates is in increasing order.

\begin{observation}\label[observation]{obs:repr-linearity}
For every integer $N\ge 1$ and vectors $\vec x,\vec y\in \mathbb N^k$,
$
\repr_N(\vec x+\vec y)=\repr_N(\vec x)+\repr_N(\vec y).
$
\end{observation}

\begin{proof}
By definition,
\[
\repr_N(\vec x+\vec y)
=
\sum_{b=0}^{k-1}(x_b+y_b)N^b
=
\sum_{b=0}^{k-1}x_bN^b+\sum_{b=0}^{k-1}y_bN^b
=
\repr_N(\vec x)+\repr_N(\vec y).
\]
\end{proof}

\paragraph{$3$-term arithmetic progressions.}
A $3$-term arithmetic progression ($3$-AP) is a sequence of the form
$
a,a+d,a+2d,
$
where $a,d\in \mathbb{Z}$. 
The $3$-AP is called \emph{nontrivial} if $d\neq 0$. For an integer $N\ge 1$, let $r_3(N)$ denote the maximum size of a subset of $\{1,\dots,N\}$ containing no nontrivial $3$-term arithmetic progression.
Both upper and lower bounds for $r_3(N)$ have been extensively studied in the literature~\cite{roth1953certain, heath1987integer, szemeredi1990integer, bourgain1999triples, bourgain2008roth, sanders2011roth, bloom2016quantitative, schoen2021improved, bloom2020breaking, kelley2023strong, bloom2023kelley, raghavan2026improved, erdos1936some, salem1942sets, behrend1946sets, elkin2010improved, o2008sets, elsholtz2024improving}.
Since our paper proves lower bounds, we will mostly use lower bounds on $r_3(N)$.
The state-of-the-art bounds for $r_3(N)$ are summarized in the following theorem.

\begin{theorem}[~\cite{behrend1946sets, elsholtz2024improving, raghavan2026improved}]
\label{thm:r3-known-bounds}
There exist constants\footnote{The lower bound was already shown in the 1940s  by Behrend~\cite{behrend1946sets} who proved the bound for $c_1= 2\sqrt 2+o(1)$. 

The current best known lower bound, due to Elsholtz, Hunter, Proske, and Sauermann~\cite{elsholtz2024improving}, improves this constant to $c_1 = 2\sqrt{\log_2(24/7)}+o(1)$.
} $c_1,c_2>0$ such that, for all sufficiently large $N$, 
\[ \frac{N}{2^{c_1\sqrt{\log N}}}\le r_3(N)\le \frac{N}{2^{c_2(\log N)^{1/6}/\log\log N}}. \]
\end{theorem}

\section{Layered Graphs from 3-AP-Free Sets}\label{sec:construction}

In this section we describe the construction of the graph which we use in \cref{sec:generalLBs} to prove \cref{thm:distance-labeling-sumindex}. 
Our construction is inspired by the construction of Berman, Raskhodnikova and Ruan~\cite{BermanRR10}. 
However, their construction uses extremal points of convex hulls, while we use 3-AP-free sequences, which turn out to be a better fit for the Sum-Index problem.

\begin{lemma}\label[lemma]{thm:avgfree-layered-CP}
For all integers $N,k\ge 1$  and every $3$-AP-free set $A\subseteq \lrb N$, there exists an unweighted undirected graph $G$, constructed for tuple $(N,k,A)$ with the following properties.
\begin{enumerate}
    \item The graph has \(2k+1\) layers, each indexed by \(\lbb{3N}\rbb^k\).
Thus, the graph has \(n=(2k+1)(3N)^k\) vertices.

    \item The graph has $m
    =O\!\left(n\cdot |A|\right)=O(n^{1+1/k})$ edges.

    \item There is a set \(\mathcal P=\{P_{\vec u,\vec s}\mid
\vec u\in \lbb N\rbb^k,\ \vec s\in A^k\}\) of
$|\mathcal P|=N^k|A|^k
=
\frac{n|A|^k}{(2k+1)3^k}$
critical paths. For \(\vec v=\vec u+2\vec s\), the path
\(P_{\vec u,\vec s}\) is the unique shortest path of length \(2k\)
from \((0,\vec u)\) to \((2k,\vec v)\), and it passes through
\((k,\vec u+\vec s)=(k,(\vec u+\vec v)/2)\).
\end{enumerate}
\end{lemma}

\begin{proof}
We construct the graph $G=(V,E)$ as follows.

\paragraph{Vertices and edges.} The vertices are defined in the statement. 
We denote each vertex in the graph as $(\ell, \vec u)$, 
where $\ell\in \lbb{2k+1}\rbb$ is the layer index and $\vec u\in \lrb{3N}^k$.

For every vertex $(\ell,\vec u)\in V$ 
where $\ell\in \lrb {2k}$, let $b=\ell \bmod k$. 
We add edges from $(\ell,\vec u)$ to every vertex of the form 
\[
(\ell+1,(u_0,\dots,u_{b-1},u_b+s,u_{b+1},\dots,u_{k-1})), \qquad \text{where} \,\,s\in A \text{\, and } (u_b+s)\in \lrb{3N}  .
\]
Hence every vertex has at most 
$|A|$
edges connecting to the next layer, and therefore
$$m\le2k(3N)^k\cdot |A|=O\!\left(n\cdot |A|\right)=O(nN)=O(n^{1+1/k}).$$

\paragraph{Critical paths.}
For every \(\vec u\in \lbb N\rbb^k\) and every
\(\vec s=(s_0,\ldots,s_{k-1})\in A^k\), define \(P_{\vec u,\vec s}\)
as follows. 
Starting from \((0,\vec u)\), at layer
\(\ell < 2k\), the path
uses the edge that increases coordinate \(b = \ell \bmod k\) by \(s_b\).
Since each coordinate $b$ is increased twice (once at layer \(b\) and once
at layer \(k+b\)) and
\(u_b+2s_b<3N\), the path is well defined. It has length \(2k\),
ends at \((2k,\vec u+2\vec s)\), and passes through
\((k,\vec u+\vec s)\).

We now prove uniqueness. Any path from layer \(0\) to layer \(2k\)
has length at least \(2k\). Consider a path \(P'\) of length \(2k\)
from \((0,\vec u)\) to \((2k,\vec u+2\vec s)\). Since \(P'\) has
minimum possible length, it uses exactly one edge between each pair of
consecutive layers. Hence, for every coordinate \(b\), the path
increases coordinate \(b\) exactly twice. Let these increases be \(x_b,y_b\in A\). Since the
endpoint is \(\vec u+2\vec s\), we have
$
x_b+y_b=2s_b.
$
As \(A\) is 3-AP-free and \(x_b,s_b,y_b\in A\), we have
\(x_b=y_b=s_b\). 
Thus \(P'\) uses the same increases as
\(P_{\vec u,\vec s}\) at every layer, and therefore \(P'=P_{\vec u,\vec s}\).

The number of critical paths is \(N^k|A|^k\), one for each choice of
\(\vec u\in\lbb N\rbb^k\) and \(\vec s\in A^k\), which equals
\(\frac{n|A|^k}{(2k+1)3^k}\).

\end{proof}

\begingroup
\setlength{\intextsep}{6pt}
\setlength{\abovecaptionskip}{4pt}
\setlength{\belowcaptionskip}{0pt}
\begin{figure}[!htbp]
  \centering
  \resizebox{0.9\linewidth}{!}{%
\begin{tikzpicture}[
  x={(1.80cm,0cm)}, y={(.25cm,.12cm)}, z={(0cm,1.20cm)},
  font=\small,
  cp/edge/.style={draw=cpSlate!26,line width=.28pt},
  cp/grid/.style={draw=cpSlate!25,line width=.22pt,densely dotted},
  cp/first/.style={draw=cpBlue,line width=1.3pt,
    preaction={draw=white,line width=2.2pt}},
  cp/second/.style={draw=cpRed,line width=1.3pt,
    preaction={draw=white,line width=2.2pt}},
  cp/label/.style={anchor=east,font=\small,inner sep=2pt},
  cp/card/.style={anchor=west,align=left,font=\scriptsize,
    text width=25mm,rounded corners=2pt,inner sep=5pt,
    draw=cpSlate!35,fill=cpSlate!3,line width=.4pt},
]
\definecolor{cpBlue}{HTML}{0072B2}
\definecolor{cpRed}{HTML}{D1493F}
\definecolor{cpTeal}{HTML}{00836B}
\definecolor{cpSlate}{HTML}{4C6275}

\foreach \ell in {0,...,4}{
  \ifnum\ell=2
    \filldraw[fill=cpTeal!5,draw=cpTeal!35,line width=.5pt]
      (-.3,-.3,\ell) -- (3.3,-.3,\ell) --
      (3.3,6.3,\ell) -- (-.3,6.3,\ell) -- cycle;
  \else
    \filldraw[fill=cpSlate!3,draw=cpSlate!22,line width=.4pt]
      (-.3,-.3,\ell) -- (3.3,-.3,\ell) --
      (3.3,6.3,\ell) -- (-.3,6.3,\ell) -- cycle;
  \fi
  \foreach \xx in {0,...,3}{
    \draw[cp/grid] (\xx,0,\ell) -- (\xx,6,\ell);
  }
  \foreach \yy in {0,...,6}{
    \draw[cp/grid] (0,\yy,\ell) -- (3,\yy,\ell);
  }
}

\foreach \ell in {0,...,3}{
  \pgfmathtruncatemacro{\nextlayer}{\ell+1}
  \foreach \a in {0,1,3}{
    \ifodd\ell
      \pgfmathtruncatemacro{\lastsource}{6-\a}
      \foreach \xx in {0,...,3}{
        \foreach \yy in {0,...,\lastsource}{
          \pgfmathtruncatemacro{\nextcoord}{\yy+\a}
          \draw[cp/edge] (\xx,\yy,\ell) -- (\xx,\nextcoord,\nextlayer);
        }
      }
    \else
      \pgfmathtruncatemacro{\lastsource}{3-\a}
      \foreach \xx in {0,...,\lastsource}{
        \pgfmathtruncatemacro{\nextcoord}{\xx+\a}
        \foreach \yy in {0,...,6}{
          \draw[cp/edge] (\xx,\yy,\ell) -- (\nextcoord,\yy,\nextlayer);
        }
      }
    \fi
  }
}

\coordinate (cp0) at (0,0,0);
\coordinate (cp1) at (1,0,1);
\coordinate (cp2) at (1,3,2);
\coordinate (cp3) at (2,3,3);
\coordinate (cp4) at (2,6,4);
\draw[cp/first] (cp0) -- (cp1);
\draw[cp/second] (cp1) -- (cp2);
\draw[cp/first] (cp2) -- (cp3);
\draw[cp/second] (cp3) -- (cp4);
\foreach \ell in {0,...,4}{
  \path[fill=white,draw=cpSlate!78,line width=.25pt]
    \foreach \xx in {0,...,3}{
      \foreach \yy in {0,...,6}{
        (\xx,\yy,\ell) circle[radius=1.05pt]
      }
    };
}
\foreach \p in {cp0,cp1,cp3,cp4}{
  \filldraw[fill=cpBlue,draw=white,line width=.75pt] (\p) circle[radius=2.15pt];
}
\filldraw[fill=cpTeal,draw=white,line width=.85pt] (cp2) circle[radius=2.9pt];
\draw[cpTeal,line width=.5pt] (cp2) circle[radius=3.9pt];

\foreach \ell/\xx/\yy/\col in {0/0/0/cpBlue,1/1/0/cpBlue,2/1/3/cpTeal,3/2/3/cpBlue,4/2/6/cpBlue}{
  \node[cp/label,text=\col] (cplabel\ell)
    at (-1.45cm,{(1.20*\ell+.12*\yy)*1cm})
    {$(\ell,(\xx,\yy))$};
  \node[anchor=south east,font=\scriptsize,text=cpSlate,inner sep=1pt]
    at ($(cplabel\ell.north east)+(0,.04cm)$) {$L_{\ell}$};
  \draw[\col!65,densely dotted,line width=.4pt]
    (cplabel\ell.east) -- ($(cp\ell)+(-.10cm,0cm)$);
}

\coordinate (cpAxisOrigin) at (-1.15cm,-.16cm);
\draw[-{Stealth[length=1.5mm]},cpSlate!80,line width=.5pt]
  (cpAxisOrigin) -- ++(5.90cm,0cm)
  node[midway,below=3pt] {$x_0$};
\draw[-{Stealth[length=1.5mm]},cpSlate!80,line width=.5pt]
  (cpAxisOrigin) -- ++(0,4,0)
  node[pos=.55,above=3pt,sloped,xshift=3pt] {$x_1$};
\node[font=\scriptsize,text=cpSlate,below left=2pt,xshift=4pt]
  at (cpAxisOrigin) {$(0,0)$};

\node[cp/card] at (8.00cm,2.85cm) {
  $k=2,\quad N=4$\par
  $A=\{0,1,3\}$\par
  $\vec s=(1,3)$\par
  \smallskip
  $\vec u=(0,0)$\par
  $\vec v=(2,6)$\par\medskip
  \begin{tabular}{@{}c@{\hspace{4pt}}l@{}}
    \tikz[baseline=-.55ex]{\draw[cpBlue,line width=1.1pt] (0,0)--(.35cm,0);}
      & $x_0\gets x_0+1$\\[2pt]
    \tikz[baseline=-.55ex]{\draw[cpRed,line width=1.1pt] (0,0)--(.35cm,0);}
      & $x_1\gets x_1+3$\\[2pt]
    \tikz[baseline=-.55ex]{\filldraw[fill=cpTeal,draw=white,line width=.6pt] (0,0) circle[radius=1.8pt];
      \draw[cpTeal,line width=.4pt] (0,0) circle[radius=2.6pt];}
      & $(\vec u+\vec v)/2$
  \end{tabular}
};
\end{tikzpicture}
}%
  \caption{An example of the graph construction and a critical path from Lemma~\ref{thm:avgfree-layered-CP}, for $k=2$, $N=4$, and the
  $3$-AP-free set $A=\{0,1,3\}$.
  Layers $L_0,\ldots,L_4$ appear from bottom to top. 
  The highlighted critical path \(P_{(0,0),(1,3)}\) has endpoints
  $(0,\vec u)$ and $(4,\vec v)$, where $\vec u=(0,0)$,
  $\vec s=(1,3)$, and $\vec v=\vec u+2\vec s=(2,6)$.
  Blue edges increment the $x_0$ coordinate by $1$, and red edges increment the $x_1$ coordinate
  by $3$.
  The green midpoint is $(2,(1,3))$, with coordinate vector
  $(1,3)=\vec u+\vec s=(\vec u+\vec v)/2$.
  The illustration is truncated: only vertices with coordinate vectors in
  $\{0,\ldots,3\}\times\{0,\ldots,6\}$ are displayed in each layer,
  instead of the full set $\{0,\ldots,11\}^2$.
  All graph edges are undirected and unweighted.}
  \label{fig:ap-free-critical-path}
\end{figure}
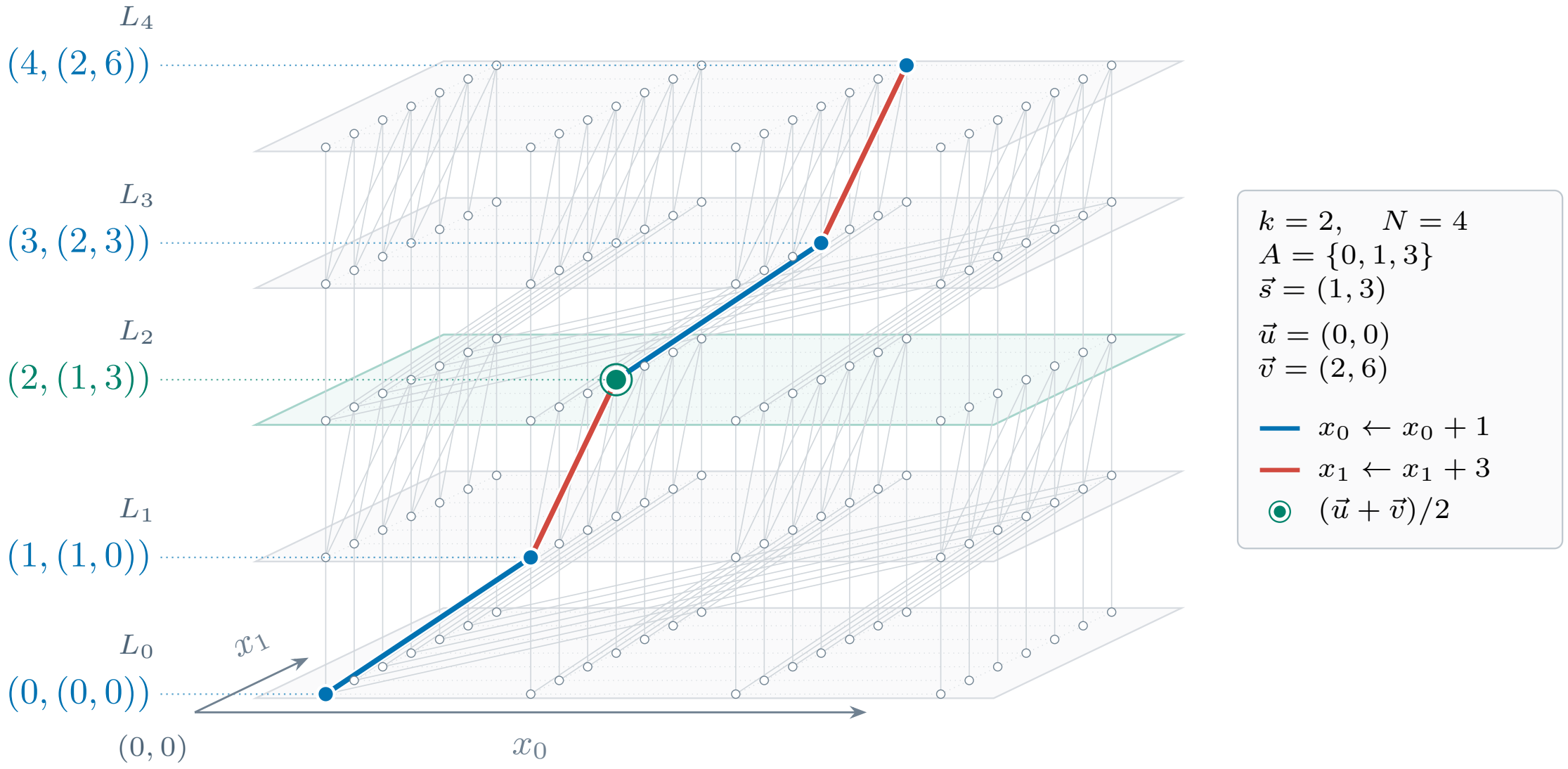
\endgroup

\section{Reductions Between Sum-Index Variants}\label{sec:sumindexreduc}
To prove our new lower bounds we reduce to a  variation of the Sum-Index problem. 
In this section, we prove a reduction chain between Sum-Index variations which terminates with a variation that better fits our graph construction from \cref{sec:construction}.

The lower-bound proofs use the following two results about the Sum-Index problem.

\begin{theorem}[Sum-Index lower bound~\cite{nisan1993rounds,babai1995simultaneous,pudlak1997boolean,babai2003communication}]
\label{thm:sumindex-lower-bound}
For every positive integer $n$,
$
\SI(n)=\Omega(\sqrt n).
$
\end{theorem}

\begin{lemma}\label[lemma]{thm:sumindex-constant-factor}
For any integer parameters $n\ge k \ge 1 $, we have $\SI(kn)=O(k\cdot\SI(n))$.
\end{lemma}
A proof for \cref{thm:sumindex-constant-factor} is included in \cref{appndx:missingProofs}.

\subsection{Restricted Sum-Index}
The first variation we consider is the $D$-restricted Sum-Index problem.

\begin{definition}
For an integer $n\ge 1$ and $D \subseteq \lrb n$, the \emph{$D$-restricted Sum-Index problem} is defined in the same way as the Sum-Index problem, except that 
the inputs $i$ and $j$ are promised to satisfy 
\[
(j-i)\bmod n \in D.
\]
The communication complexity of the $D$-restricted Sum-Index problem is denoted by $\SI_D(n)$.
\end{definition}

Notice that $\SI(n)=\SI_{\lrb{n}}(n)$. The following lemma relates $\SI(n)$ and $\SI_D(n)$.

\begin{lemma}\label[lemma]{lem:restricted-sum-index-reduction}
For every integer $n\ge 2$ and nonempty set $D\subseteq \lrb n$,
\[
\SI(n)
\le
O\!\left(\frac{n\log n}{|D|}\right)\cdot \SI_D(n).
\]
\end{lemma}

\medskip

The proof of \cref{lem:restricted-sum-index-reduction} relies on the following lemma.

\begin{lemma}\label[lemma]{lem:small-shift-cover}
For any integer $n\ge 2$ and nonempty set $D\subseteq \lrb n$, there exists a set $T\subseteq \lrb n$ of size $|T|=O\!\left(\frac{n\log n}{|D|}\right)$ 
such that $$
(D+T)\bmod n := \{(d+t)\bmod n \mid d\in D,\ t\in T\}= \lrb n.
$$

\end{lemma}

\begin{proof}
For a fixed  $x\in \lrb n$ and uniformly random $t\in \lrb n$, 
 $\Pr[x\in 
(D+\{t\})\bmod n] = |D|/n$. 
Let $T$ be a multiset of $m:=\left\lceil \frac{2n\ln n}{|D|}\right\rceil$ elements chosen (with repetition) independently and uniformly at random from $\lrb n$.
By a hitting-set argument and  union bound, $$\Pr[((D+T)\bmod n) \neq \lrb n]\le n\cdot
 \left(1-\frac{|D|}{n}\right)^m
 \le n\cdot e^{-m|D|/n}
 \le n\cdot e^{-2\ln n}
 = n\cdot  n^{-2}  <1,$$
and so there exists a set $T$ (after removing duplicates) of size at most $m=O(n\log n/|D|)$ such that $(D+T)\bmod n=\lrb n$.
\end{proof}

We are now ready to prove \cref{lem:restricted-sum-index-reduction}.

\begin{proof}[Proof of \cref{lem:restricted-sum-index-reduction}]
Recall that $S$ is the common input string, and  Alice and Bob receive indices $i,j\in \lrb n$, respectively.

By \cref{lem:small-shift-cover}, there exists a set $T$ of size $|T|=O\!\left(\frac{n\log n}{|D|}\right)$ such that $(D+T)\bmod n=\lrb n$.
Fix any such set $T$.
For every $x,y\in \lrb n$, define $S^{(-x)}[y]= S[(y-x)\bmod n]$, which is a cyclic shift of $S$.

Let $\Pi_D$ be a protocol for the $D$-restricted Sum-Index problem with communication complexity \(\SI_D(n)\). 
We construct a protocol for Sum-Index as follows.
First, Alice and Bob send $i$ and $j$, respectively, to the referee.
Next, for every \(t\in T\), Alice sends the message prescribed by
\(\Pi_D\) on common string \(S^{(-t)}\) and index \((i+t)\bmod n\),
while Bob sends the message prescribed by \(\Pi_D\) on common string
\(S^{(-t)}\) and index \(j\).

The referee  computes $\delta := (j-i)\bmod n$.
By \cref{lem:small-shift-cover}, there exists $t\in T$ such that  
 $(\delta-t)\bmod n \in D$.
 The referee chooses such a value $t$, and performs the referee's decision procedure on the messages from Alice and Bob that were sent during their  execution of $\Pi_D$ on common string $S^{(-t)}$ and indices $(i+t)\bmod n$ and $j$, respectively.
 Notice that the communication overhead required for the referee to identify which messages correspond to the chosen $t$ is avoided by fixing an ordering of the elements of $T$.
 
Since
\[
(j-((i+t)\bmod n))\bmod n
=
(j-i-t)\bmod n
=
(\delta-t)\bmod n
\in D,
\]
the indices $(i+t)\bmod n$ and $j$ satisfy the promise of the $D$-restricted Sum-Index problem. 
Thus, the referee recovers the value of 
 \[
S^{(-t)}[((i+t)+j)\bmod n]
=
S[(((i+t)+j)-t)\bmod n]
=
S[(i+j)\bmod n].
\]

For the communication analysis, protocol $\Pi_D$ is executed $O(n\log n / |D|)$ times, and, in addition, Alice and Bob each send another $O(\log n)$ bits for indices $i$ and $j$, for a total of 
\[
|T|\cdot \SI_D(n)+O(\log n)
=
O\!\left(\frac{n\log n}{|D|}\right)\cdot \SI_D(n),
\]
where the additive $O(\log n)$ term is absorbed since $\log n\le \frac{n\log n}{|D|}$ and $1 \le \SI_D(n)$.
\end{proof}

\subsection{Vector Restricted Sum-Index}
In the next variant of the Sum-Index problem we reduce to, Alice and Bob receive vectors instead of numbers.

\begin{definition}
Let $\N,k\ge 1$ be integers, and let $D\subseteq \lbb \N \rbb^k$. The \emph{vector $D$-restricted Sum-Index problem} is defined in the same way as the Sum-Index problem, except that the common input string has length $\N^k$, Alice and Bob receive vectors $\vec{i},\vec{j}\in \lbb {\N}\rbb^k$ instead of indices in $\lrb{{\N}^k}$, and the inputs are promised to satisfy
\[
\vec{j}-\vec{i}\in D
\qquad\text{and}\qquad
\vec{i}+\vec{j}\in \lbb {\N}\rbb^k.
\]
The referee must output
\[
S[\repr_{\N}(\vec{i}+\vec{j})].
\]
The communication complexity of the vector $D$-restricted Sum-Index problem is denoted by $\VSI_D({\N},k)$.
\end{definition}

We reduce the $D$-restricted Sum-Index problem to the vector $D$-restricted Sum-Index problem. 
The reduction uses the base ${\N}$ representation of indices, and so  addition of indices becomes coordinatewise addition of digit vectors, up to carries.
Notice that in the definition of the vector $D$-restricted Sum-Index problem, we add the condition $\vec{i}+\vec{j}\in \lbb {\N}\rbb^k$, which does not have a corresponding condition in the previous Sum-Index variations. 
This is so we do not face modularity issues in the vector coordinates.
Our reduction overcomes this relaxation by allowing an enlarged coordinate universe of $3N$ instead of $N$.

\begin{lemma}\label[lemma]{lem:restricted-to-vector-sum-index}
For all integers ${\N}\ge 2, k\ge 1$ and every nonempty set $D\subseteq \lbb {\N}\rbb^k$, let 
\[
D^{\mathrm{num}}:=\{\repr_{\N}(\vec{d})\mid \vec{d}\in D\}\subseteq \lrb{{\N}^k}.
\]
Then
\[
\SI_{D^{\mathrm{num}}}({\N}^k)\le O(2^k)\cdot \VSI_D(3{\N},k)
+O(k\log {\N})
.
\]
\end{lemma}

\begin{proof}
Notice that \(D\) is  a subset of \(\lbb{3{\N}}\rbb^k\), and so $\VSI_D(3{\N},k)$ is well defined. 
Let $\Pi_{\mathrm{vec}}$ be a protocol for the vector $D$-restricted Sum-Index problem with communication complexity $\VSI_D(3{\N},k)$.
We construct a protocol for the $D^{\mathrm{num}}$-restricted Sum-Index problem on strings of length ${\N}^k$.
Let $S\in\{0,1\}^{{\N}^k}$ be the common input string, and let Alice and Bob receive indices $i,j\in \lrb{{\N}^k}$,
respectively, promised to satisfy 
\[
(j-i)\bmod {\N}^k\in D^{\mathrm{num}}.
\]
Let \(d:=(j-i)\bmod \N^k\). Since \(d\in D^{\mathrm{num}}\), there is a unique \(\vec d\in D\) such that \(d=\repr_\N(\vec d)\). Hence
\begin{equation}\label{eq:sum-mod-Nk}
j=(i+d)\bmod \N^k.
\end{equation}

Let $\vec i$, $\vec j$ be the base ${\N}$ representations of $i$ and $j$, respectively, so $i=\repr_{\N}(\vec{i})$ and $j=\repr_{\N}(\vec{j})$, 
and let
\[
\vec{i}=(i_0,\dots,i_{k-1}),\qquad
\vec{j}=(j_0,\dots,j_{k-1}),\qquad
\vec{d}=(d_0,\dots,d_{k-1})\qquad\in \lbb {\N}\rbb^k.
\]
Equation~\eqref{eq:sum-mod-Nk} corresponds to digitwise addition with carries. 
Let $c_{-1}=0$, and let $\vec{c}:=(c_0,\dots,c_{k-1})\in\{0,1\}^{k}$, where for every $b\in \lbb k\rbb$, $c_b=
1$ if and only if $i_b+d_b+c_{b-1}\ge {\N}$.
Thus, 
for every $b\in \lbb k\rbb$, 
\begin{equation}\label{eq:carry}
j_b=i_b+d_b+c_{b-1}-{\N}c_b,
\end{equation}
with the last carry $c_{k-1}$ accounting for the reduction modulo ${\N}^k$.

For each guessed carry vector $\vec \gamma \in \{0,1\}^k$, we define one simulated vector instance over the universe $\lbb 3{\N}\rbb ^k$.
Let $\gamma_{-1}=0$.
For every $\vec \gamma \in \{0,1\}^k$, Alice defines a vector $\vec{i}^{\,\vec{\gamma}}\in \lbb{3{\N}}\rbb^k$ by
\[
i^{\,\vec{\gamma}}_b:=i_b+\gamma_{b-1},
\qquad\text{for every } b\in \lbb k\rbb,
\]
and Bob defines a vector $\vec{j}^{\,\vec{\gamma}}\in \lbb{3{\N}}\rbb^k$ by
\[
j^{\,\vec{\gamma}}_b:=j_b+{\N}\gamma_b,
\qquad\text{for every } b\in \lbb k\rbb.
\]
In addition, define a string $S_{\vec{\gamma}}\in\{0,1\}^{(3{\N})^k}$ as follows. For every $\vec{x}=(x_0,\dots,x_{k-1})\in \lbb{3{\N}}\rbb^k$, let
\[
S_{\vec{\gamma}}[\repr_{3{\N}}(\vec{x})]
:=
S\!\left[\left(\sum_{b=0}^{k-1}(x_b-\gamma_{b-1}-{\N}\gamma_b){\N}^b\right)\bmod {\N}^k\right].
\]
Thus, for every $\vec{\gamma}\in \{0,1\}^k$ and every $b\in \lbb k \rbb$ we have $(i^{\vec{\gamma}}_b + j^{\vec{\gamma}}_b)-\gamma_{b-1}-{\N}{\gamma_b} = i_b+j_b$, and so  
\begin{equation}\label{eq:S_gamma}
S_{\vec\gamma}[\repr_{3{\N}}(\vec{i}^{\,\vec\gamma}+\vec{j}^{\,\vec\gamma})]
=
S\!\left[\left(\sum_{b=0}^{k-1}(i_b+j_b)\N^b\right)\bmod \N^k\right]
=
S[(i+j)\bmod \N^k].
\end{equation}

For every $\vec \gamma \in \{0,1\}^k$, in a fixed canonical order, Alice and Bob send the messages prescribed by $\Pi_{\mathrm{vec}}$ on common string $S_{\vec{\gamma}}$, and private inputs $\vec{i}^{\,\vec{\gamma}}$ and $\vec{j}^{\,\vec{\gamma}}$, respectively. 
Finally, Alice sends $i$ and Bob sends $j$ to the referee.

Using \(i\) and \(j\), the referee computes \(\vec i,\vec j\),
\(d=(j-i)\bmod {\N}^k\), the unique \(\vec d\in D\) satisfying
\(\repr_{\N}(\vec d)=d\), and the carry vector \(\vec c\)
defined by  $c_{-1} = 0$ and $c_b = 1$ if and only if $i_b+d_b+c_{b-1} \ge {\N}$.
The referee uses  $\vec c$ to identify the messages sent by Alice and Bob during the execution of $\Pi_{\mathrm{vec}}$ for $\vec{\gamma}=\vec{c}$.

By the definition of the vector $D$-restricted Sum-Index problem, to justify that the referee can recover the value of $S_{\vec{c}}[\repr_{3{\N}}(\vec{i}^{\,\vec{c}}+\vec{j}^{\,\vec{c}})]$, we need to verify that 
\[
\vec{j}^{\,\vec{c}}-\vec{i}^{\,\vec{c}}\in D
\qquad\text{and}\qquad
\vec{i}^{\,\vec{c}}+\vec{j}^{\,\vec{c}}\in \lbb{3{\N}}\rbb^k.
\]
For the first condition, using Equation~\eqref{eq:carry}, we have that for every $b\in \lbb k\rbb$,
\[
j^{\,\vec{c}}_b=j_b+{\N}c_b=i_b+d_b+c_{b-1}=i^{\,\vec{c}}_b+d_b, 
\]
and so $\vec{j}^{\,\vec{c}}-\vec{i}^{\,\vec{c}}=\vec{d}\in D$.
For the second condition, we have that for every $b\in \lbb k\rbb$,
\[
i^{\,\vec{c}}_b+j^{\,\vec{c}}_b
=
(i_b+c_{b-1})+(j_b+{\N}c_b)
\le
({\N}-1)+1+({\N}-1)+{\N}
=
3{\N}-1,
\]
and so $\vec{i}^{\,\vec{c}}+\vec{j}^{\,\vec{c}}\in \lbb{3{\N}}\rbb^k$.
Thus,  the referee recovers $S_{\vec{c}}[\repr_{3{\N}}(\vec{i}^{\,\vec{c}}+\vec{j}^{\,\vec{c}})]$,
which by Equation~\eqref{eq:S_gamma} is the desired output
$
S[(i+j)\bmod {\N}^k]
$.

For the communication analysis, there are $2^k$ possible carry vectors, and for each carry vector, each player sends the messages corresponding to one simulation of the protocol $\Pi_{\mathrm{vec}}$. 
The messages of the simulated protocol are sent in an arbitrary canonical ordering of the vectors $\vec \gamma \in \{0,1\}^k$, so no extra communication is needed.
Finally, the players send $i$ and $j$, which requires $O(\log({\N}^k))=O(k\log {\N})$ bits. Therefore the total communication is
\[
O(2^k)\cdot \VSI_D(3{\N},k)+O(k\log {\N}),
\]
which proves the lemma.
\end{proof}

The following corollary summarizes the reduction chain we proved in this section.

\begin{corollary}\label[corollary]{cor:sumindex-to-vector}
Let ${\N}\ge 2, k\ge 1$ and $n={\N}^k$ be integers.
Let $C>0$ be a sufficiently large constant, and let
$D\subseteq \lbb {\N}\rbb^k$ such that
$|D|\ge C\sqrt n(\log n)^2$.
Then,
\[
\SI(n)
\le
O\!\left(
\frac{2^k n\log n}{|D|}
\right)\cdot
\VSI_D(3{\N},k).
\]
\end{corollary}

\begin{proof}
Let
$
D^{\mathrm{num}}:=\{\repr_{\N}(\vec d)\mid \vec d\in D\}\subseteq \lrb n$.
Since every vector $\vec d\in D\subseteq \lbb {\N}\rbb^k$ has a different base ${\N}$ representation, we have $|D^{\mathrm{num}}|=|D|$.
By \cref{lem:restricted-sum-index-reduction} and \cref{lem:restricted-to-vector-sum-index}, and since $k\log {\N}=\log n$, we have
\[
\SI(n)
\le
O\!\left(\frac{n\log n}{|D|}\right)
\cdot
\left(
O(2^k)\cdot \VSI_D(3{\N},k)+O(\log n)
\right).
\]
Using the fact that $|D|\ge C\sqrt n(\log n)^2$, we bound the contribution of the additive $O(\log n)$ term in the last equation:
\[
O\!\left(\frac{n\log n}{|D|}\cdot \log n\right)
\le
O\!\left(\frac{n(\log n)^2}{C\sqrt n(\log n)^2}\right)
=
O\!\left(\frac{\sqrt n}{C}\right).
\]
By \cref{thm:sumindex-lower-bound}, $\SI(n)=\Omega(\sqrt n)$. Thus, by choosing $C$ sufficiently large, the term $O(\sqrt{n}/C)$ is dominated by $\SI(n)$. 
Therefore, 
$
\SI(n)
\le
O\!\left(\frac{2^k n\log n}{|D|}\right)\cdot
\VSI_D(3{\N},k).
$\end{proof}

\section{Approximate Distance Labeling Lower Bound}\label{sec:generalLBs}

Given the graph construction in \cref{sec:construction} and the reduction from Sum-Index to $D$-restricted Sum-Index from \cref{sec:sumindexreduc}, we are now ready to prove \cref{thm:distance-labeling-sumindex}.

\begin{proof}[Proof of \cref{thm:distance-labeling-sumindex}]
Fix a $3$-AP-free set $A\subseteq \lbb N\rbb$ of size $|A|=r_3(N)$, and define
$
D:=A^k\subseteq \lbb N\rbb^k.
$

Let $G_0$ be the graph obtained by the construction of \cref{thm:avgfree-layered-CP}, constructed with the tuple $(3N,k,A)$. Thus, $G_0$ has
$
n=(2k+1)(9N)^k
$
vertices and
$
m=O\!\left(n\cdot |A|\right)
=
O\!\left(n\cdot r_3(N)\right)
=
O(n^{1+1/k})
$
edges.

For every string $X\in\{0,1\}^{(3N)^k}$, we define a graph $G_X$ on the same vertex set as $G_0$ as follows. For every vector $\vec y\in\lbb{3N}\rbb^k$, if
$
X[\repr_{3N}(\vec y)]=0,
$
then delete all edges incident to the vertex $(k,\vec y)$. The rest of the edges remain unchanged.

Fix any distance labeling scheme $\mathcal{L}$ with stretch strictly smaller than $1+1/k$ for the graph family
$
\{G_X\mid X\in\{0,1\}^{(3N)^k}\}.
$
Let $L$ be the maximum label length of this labeling scheme.
We show a lower bound for $L$ by constructing a protocol for the vector $D$-restricted Sum-Index problem with parameters $3N$ and $k$ that uses $\mathcal L$. 

Let $S\in\{0,1\}^{(3N)^k}$ be the common input string. Alice receives $\vec i\in\lbb{3N}\rbb^k$, and Bob receives $\vec j\in\lbb{3N}\rbb^k$, promised to satisfy
$\vec j-\vec i\in D$ and 
$\vec i+\vec j\in\lbb{3N}\rbb^k$.
The referee must output
$
S[\repr_{3N}(\vec i+\vec j)].
$

Let $\vec d:=\vec j-\vec i$. Thus, $\vec j\ge \vec d$ coordinatewise, and so 
\[
2\vec i=\vec i+\vec j-\vec d\in \lbb{3N}\rbb^k,
\] 
Similarly, since
$\vec d\in D=A^k\subseteq \lbb N\rbb^k$, 
\[
2\vec j=\vec i+\vec j+\vec d\in \lbb{4N}\rbb^k
\subseteq \lbb{9N}\rbb^k.
\]
Thus \(u:=(0,2\vec i)\) and \(v:=(2k,2\vec j)\) are vertices of \(G_0\).
Moreover, since we  applied \cref{thm:avgfree-layered-CP} with  tuple
\((3N,k,A)\), the vertices \(u\) and \(v\) are connected by the
critical path \(P:=P_{2\vec i,\vec d}\). This path has length \(2k\),
is the unique shortest  path from $u$ to $v$ in \(G_0\), and passes through
$w=(k,2\vec i+\vec d)=(k,\vec i+\vec j).
$

If \(S[\repr_{3N}(\vec i+\vec j)]=1\), then the edges incident to \(w\) are not deleted and
the critical path \(P\) remains, so \(\dist_{G_S}(u,v)=2k\).

Otherwise, \(S[\repr_{3N}(\vec i+\vec j)]=0\), and so  all edges incident to \(w\)
are deleted. 
Hence, the unique \(2k\)-edge  path from $u$ to $v$ in \(G_0\) is not in $G_S$. 
Since $G_S$ is layered, every path from layer $0$ to layer $2k$ has even length, and so 
$
\dist_{G_S}(u,v)\ge 2k+2.
$

Since \(S\) is common, both players know \(G_S\).
Thus, Alice sends the label of $u$ in the graph $G_S$, and Bob sends the label of $v$ in the graph $G_S$. The referee applies the decoder of the labeling scheme and obtains an estimate
$\hat d(u,v)$ satisfying\footnote{We assume $u$ and $v$ are connected in $G_S$. Otherwise, the decoder returns $\infty$, and the referee returns $0$.} $$\dist_{G_S}(u,v)\le \hat d(u,v) < (1+1/k)\dist_{G_S}(u,v).$$
If $S[\repr_{3N}(\vec i+\vec j)]=1$, then $\dist_{G_S}(u,v)=2k$, and 
$\hat d(u,v)
<
\left(1+1/k\right)2k
=
2k+2$.
Otherwise, $S[\repr_{3N}(\vec i+\vec j)]=0$, then $\dist_{G_S}(u,v)\ge 2k+2$, and 
$
\hat d(u,v)\ge 2k+2.
$
Thus, the referee correctly recovers
\[
S[\repr_{3N}(\vec i+\vec j)]
=
\begin{cases}
1, & \text{if } \hat d<2k+2,\\[1mm]
0, & \text{if } \hat d\ge 2k+2.
\end{cases}
\]
Therefore,
\[
\VSI_D(3N,k)\le L.
\]

We next apply \cref{cor:sumindex-to-vector}. 
By \cref{thm:r3-known-bounds}, $r_3(N)=N^{1-o(1)}$. 
Thus, for all sufficiently large $N$, we have
$
|D|=r_3(N)^k
=
\omega\!\left(\sqrt{N^k}(\log(N^k))^2\right),
$
and so the size condition of \cref{cor:sumindex-to-vector} holds. Applying \cref{cor:sumindex-to-vector} with $n=N^k$ we have
\[
\SI(N^k)
\le
O\!\left(
\frac{2^k N^k\log(N^k)}{|D|}
\right)\cdot
\VSI_D(3N,k)
\le
O\!\left(
\frac{2^k N^k\log(N^k)}{r_3(N)^k}
\right)\cdot L,
\]
where the last inequality follows since $\VSI_D(3N,k)\le L$ and $|D|=r_3(N)^k$. Thus,
\[
L
\ge
\Omega\!\left(
\frac{r_3(N)^k}{2^k N^k\log(N^k)}
\cdot
\SI(N^k)
\right),
\]
which proves the theorem.
\end{proof}

\subsection{Proof of \cref{thm:distance-labeling-final-bound}}\label{sec:proof_main_lb_thm}
We now prove \cref{thm:distance-labeling-final-bound} as a corollary of \cref{thm:distance-labeling-sumindex}.

\begin{proof}[Proof of \cref{thm:distance-labeling-final-bound}]
Fix a constant integer $k\ge 1$. 
By \cref{thm:distance-labeling-sumindex}, for every sufficiently large $N$, every distance labeling scheme with stretch strictly smaller than $1+1/k$ for graphs with 
$n=(2k+1)(9N)^k$ vertices requires label size at least 
\[ \Omega\!\left( \frac{r_3(N)^k}{2^kN^k\log(N^k)}\cdot \SI(N^k) \right). \] 

By \cref{thm:r3-known-bounds}, there is a constant $C>0$ such that $r_3(N)\ge N/2^{C\sqrt{\log N}}$ for all sufficiently large $N$. 
Thus,
\[ 
\frac{r_3(N)^k}{2^kN^k\log(N^k)} 
\ge \frac{(N/2^{C\sqrt{\log N}})^k}{2^kN^k\log(N^k)} 
= \frac{1}{2^k\log(N^k)\cdot 2^{Ck\sqrt{\log N}}} 
\ge \frac{1}{2^{C'_k\sqrt{\log N}}}, \] 
for some constant $C'_k>0$ and all sufficiently large $N$. 
Since $k$ is a constant, we have $n=(2k+1)(9N)^k = \Theta_k(N^k)$, and so by \cref{thm:sumindex-constant-factor}, $\SI(N^k)=\Omega_k(\SI(n))$. Absorbing constant factors into the exponent, the required label size is at least 
\[ 
\Omega\!\left( \frac{r_3(N)^k}{2^kN^k\log(N^k)}\cdot \SI(N^k) \right)
\ge
\Omega\!\left( \frac{\SI(n)}{2^{C'_k\sqrt{\log N}}}\right)
\ge
\frac{\SI(n)}{2^{C_k\sqrt{\log n}}} \] 
for some constant $C_k>0$, completing the proof.
\end{proof}

\section{Consequences for 2-Hop Structures}\label{sec:consequences}
We present additional applications of our new graph construction from \cref{sec:construction} for $2$-hop structures. We first introduce additional preliminaries.

\subsection{Preliminaries: Shortcut Sets, Hopsets, and Hub Labels}\label{sec:extraPreliminaries}

The transitive closure of a graph $G=(V,E)$ is the graph $G^*$ with vertex
set $V$ and edge set
\[
E(G^*)=\{(u,v)\in V^2 \mid \text{there exists a path } u\leadsto v
\text{ in } G\}.
\]
The reachability diameter of a directed graph $G$ is
$\max_{(u,v)\in E(G^*)}\dist_G(u,v)$. 
Throughout, we refer to the reachability diameter simply as
the diameter.

\paragraph{Shortcut set.}
Given a directed graph $G=(V,E)$, a \emph{shortcut set with diameter} $D$ is a set
of edges $H\subseteq E(G^*)$ such that for every $(u,v)\in E(G^*)$,
$\dist_{G\cup H}(u,v)\le D$.

\paragraph{Distance hopset.}
For an integer $D\ge 1$, the $D$-hop distance
$\dist_G^D(u,v)$ is the length of the shortest path from $u$ to $v$ in $G$
that uses at most $D$ edges. If no such path exists, then
$\dist_G^D(u,v)=\infty$.

Given a weighted graph $G=(V,E)$, a \emph{hopset} with stretch $t\ge1$ and hopbound
$D$ is a set of edges $H\subseteq E(G^*)$, where every edge $(x,y)\in H$ has
weight $w(x,y)=\dist_G(x,y)$, such that for every $(u,v)\in E(G^*)$,
$\dist_G(u,v)\le \dist_{G\cup H}^D(u,v) \le t\cdot \dist_G(u,v)$.

\paragraph{Hub labeling, or $2$-hop labeling.}
In a hub labeling scheme for a directed graph $G$, the label $\ell(v)$ for $v\in V$  consists of two hub sets
$S_{\out}(v),S_{\inn}(v)\subseteq V(G)$, where each hub
$h\in S_{\out}(v)$ is stored together with $\dist_G(v,h)$, and each hub
$h\in S_{\inn}(v)$ is stored together with $\dist_G(h,v)$.

Given the labels of an ordered pair $u,v\in V(G)$, the decoder returns
\[
\hat d(u,v)
=
\min_{h\in S_{\out}(u)\cap S_{\inn}(v)}
\left(
\dist_G(u,h)+\dist_G(h,v)
\right),
\]
where $\hat d(u,v)=\infty$ if
$S_{\out}(u)\cap S_{\inn}(v)=\emptyset$. The scheme has stretch $t$ if
$\dist_G(u,v)\le \hat d(u,v)\le t\cdot \dist_G(u,v)$ 
for every ordered pair $u,v$.

For undirected graphs, the definition is analogous, except that each vertex
$v$ stores a single hub set $S(v)=S_{\out}(v)=S_{\inn}(v)$, and so for any
$u,v$, $\hat d(u,v)=\hat d(v,u)$. 
A reachability hub labeling scheme
is the analogous notion for directed reachability, where the decoder returns
whether $S_{\out}(u)\cap S_{\inn}(v)\ne\emptyset$.

The total size of a directed hub labeling scheme is
\[
\max_{G\in \mathcal G}
\sum_{v\in V(G)}
\left(
|S_{\out}(v)|+|S_{\inn}(v)|
\right),
\]
and in the undirected case it is
$\max_{G\in \mathcal G}\sum_{v\in V(G)}|S(v)|$. This size measure counts
stored hubs rather than label bit lengths.

\subsection{Equivalence Between 2-Hop Labels and 2-Hop Shortcut Sets}
We now show that hub labeling schemes are an essentially equivalent object to shortcut sets or hopsets (depending on whether the graph is weighted or not).

\begin{theorem}
\label[theorem]{lem:two-hop-labels-hopsets-equivalence}
Fix one of the following settings: weighted directed exact distances,
weighted undirected exact distances, unweighted directed exact distances,
unweighted undirected exact distances, or directed reachability. In the
distance settings, let $H$ denote an exact distance hopset; in the reachability
setting, let $H$ denote a shortcut set.

Given a graph $G=(V,E)$, a $2$-hop labeling for $G$ with total size $L$
implies a hopset or shortcut set $H$ of the same type, with $|H|\le L$ and
hopbound or diameter $2$. Conversely, a hopset or shortcut set $H$ with
hopbound or diameter $2$ implies a $2$-hop labeling of the same type with total
size $O(n+|E|+|H|)$.
\end{theorem}

\begin{proof}
The equivalence is based on the observation that every stored hub entry can be
viewed as a hopset edge, and every two-hop path through a hub gives the
corresponding hub-labeling witness.
We prove the weighted directed exact-distance case. The other cases are
identical, with the obvious changes.

Given a $2$-hop labeling, construct a hopset $H$ as follows. For every
$v\in V$ and every finite distance hub entry $x\in S_{\out}(v)$, add the edge $(v,x)$ to $H$ with
weight $\dist_G(v,x)$. For every $v\in V$ and every finite distance hub entry $x\in S_{\inn}(v)$,
add the edge $(x,v)$ to $H$ with weight $\dist_G(x,v)$. Thus, $|H|\le L$.

For every reachable pair $u,v$, correctness of the exact hub labeling implies
that the decoder returns $\dist_G(u,v)$. 
By the triangle
inequality, for every candidate hub $x$, 
$\dist_G(u,x)+\dist_G(x,v)\ge \dist_G(u,v)$, and by definition, some hub $x\in S_{\out}(u)\cap S_{\inn}(v)$ must satisfy
$\dist_G(u,v)=\dist_G(u,x)+\dist_G(x,v).
$
Therefore, $G\cup H$ contains the $2$-hop path $u\to x\to v$ of length
$\dist_G(u,v)$, and so $H$ is an exact hopset with hopbound $2$.

Conversely, suppose that $H$ is an exact hopset for $G$ with hopbound $2$.
For every edge $(u,v)\in E\cup H$, add $v$ to $S_{\out}(u)$ and $u$ to
$S_{\inn}(v)$, and in both cases store $\dist_G(u,v)$. Additionally,
for every $v\in V$, add $v$ to both $S_{\out}(v)$ and $S_{\inn}(v)$ with
distance $0$. The total size is $O(n+|E|+|H|)$.

Let $u,v$ be a reachable pair. Since $H$ has hopbound $2$, there is a path
$u\to x\to v$ in $G\cup H$ of length $\dist_G(u,v)$, where possibly $x=u$ or
$x=v$. By construction, $x\in S_{\out}(u)\cap S_{\inn}(v)$, so the decoder
returns at most $\dist_G(u,x)+\dist_G(x,v)=\dist_G(u,v)$. It cannot return a
smaller value, by the triangle inequality in $G$. Hence the decoder returns
the exact distance.
\end{proof}

We note that by applying minor changes \cref{lem:two-hop-labels-hopsets-equivalence} can be stated to also show equivalence between $t$-approximate hub labeling schemes and hopsets.

\subsection{Immediate Consequences of the Construction}\label{sec:immediate_consequences}

In addition to the maximum-label-size lower bounds for general labeling schemes from
\cref{sec:generalLBs}, our construction also implies lower bounds for
shortcut sets and for the total size of hub labeling schemes.

Similar consequences were obtained from related graph constructions in previous
work. Kosowski, Uznanski and Viennot~\cite{kosowski2019hardness} explicitly
state hub-labeling lower bounds for the undirected version of their constructed graph, and
Berman, Raskhodnikova and Ruan~\cite{BermanRR10}, whose construction inspired
\cref{sec:construction}, state related consequences in the terminology
of shortcut sets.

We first record the consequences of our new construction. For completeness, we
also state the corresponding consequences of the weighted construction from
\cite{kosowski2019hardness}, stated in Appendix~\cref{sec:KUV}. In
\cref{sec:tradeoff}, we extend these bounds to a sparsity tradeoff.

\begin{corollary}\label[corollary]{cor:2-shortcutSetLB}
For every constant integer $k\ge 2$, there are infinitely many directed
unweighted graphs $G$ with $n$ vertices, $m=O(n^{1+1/k})$ edges, and directed
diameter $2k$, such that any shortcut set $H$ that reduces the diameter of $G\cup H$
to $2$ requires size at least
\[
\Omega\!\left(n\cdot r_3(n^{1/k})^k\right)
\ge
\frac{n^2}{2^{\Theta_k(\sqrt{\log n})}}.
\]
\end{corollary}

\begin{proof}
Let $G=(V,E)$ be the directed version of the graph from
\cref{thm:avgfree-layered-CP}, applied to tuple $(N,k,A)$,
where $A\subseteq \lrb N$ is a maximum $3$-AP-free set.
That is, every edge
between consecutive layers is oriented forward. Then $G$ has $2k+1$ layers,
$n=(2k+1)(3N)^k$ vertices, and
$m=O(n\cdot r_3(N))=O(n^{1+1/k})$ edges. Since every directed path moves
forward by one layer at a time, the diameter is at most $2k$, and it
is exactly $2k$ because critical paths connect layer $0$ to layer $2k$.

Let $\mathcal P$ be the set of critical paths of $G$. We show that every
shortcut set $H$ that reduces the diameter of $G\cup H$ to $2$ has size
$\Omega(|\mathcal P|)$.

Fix a critical path $P\in\mathcal P$ from a vertex $u$ in layer $0$ to a
vertex $v$ in layer $2k$. Since $\dist_{G\cup H}(u,v)\le 2$, there is a path
from $u$ to $v$ in $G\cup H$ using at most two edges. Hence at least one of
these edges has endpoints whose layer indices differ by at least $k$. Such an
edge cannot belong to $E$, since every edge of $G$ connects consecutive layers.
We call such an edge a long shortcut for $P$.

The construction has the following prefix/suffix uniqueness property. Let $x$ be the vertex in layer $\ell$ that lies on 
critical path $P$. 
If $\ell\ge k$, then the prefix
of $P$ from layer $0$ to $x$ is contained in no other critical path. Indeed,
this prefix contains the first $k$ edges of $P$, which contain all $k$
coordinate increases of the path, and therefore determine the increase vector
$s=(s_0,\ldots,s_{k-1})$. Together with the layer-$0$ endpoint, this determines
the entire critical path.

Symmetrically, if 
$\ell\le k$, then the suffix of $P$ from $x$ to layer $2k$ is contained in no
other critical path. This suffix contains the last $k$ edges of $P$, which
again determine the increase vector $s=(s_0,\ldots,s_{k-1})$. Together with the
layer-$2k$ endpoint, this determines the entire critical path.

Since each critical path $P$ requires a long shortcut, and no two  critical paths can use the same long shortcut, we have that $|H| \ge |\mathcal{P}|$.

Finally, since $A$ is a maximum $3$-AP-free set and $k$ is a constant,  by \cref{thm:avgfree-layered-CP} we have
\[
|\mathcal P|
=
N^k |A|^k
=
\Theta\!\left(n\cdot r_3(n^{1/k})^k\right)
\ge
\frac{n^2}{2^{\Theta_k(\sqrt{\log n})}},
\]
where the final inequality follows from the known bounds for $3$-AP-free sets,
as in the proof of 
\cref{thm:distance-labeling-final-bound} (see \cref{sec:generalLBs}).
\end{proof}

The following  hub-labeling lower bound  follows from \cref{cor:2-shortcutSetLB} and the
reachability version of \cref{lem:two-hop-labels-hopsets-equivalence}.

\begin{corollary}\label[corollary]{cor:unweighted-reachability-hubLabeling}
For every constant integer $k\ge 2$, there are infinitely many directed
unweighted graphs $G$ with $n$ vertices, $m=O(n^{1+1/k})$ edges, and directed
diameter $2k$, such that any $2$-hop reachability labeling scheme for $G$
requires total size at least
\[
\Omega\!\left(n\cdot r_3(n^{1/k})^k\right)
\ge
\frac{n^2}{2^{\Theta_k(\sqrt{\log n})}}.
\]
\end{corollary}

The same lower bound also applies to approximate $2$-hop distance labeling
schemes with stretch strictly smaller than $1+1/k$ on the corresponding
unweighted undirected layered graphs with $2k+1$ layers. Indeed, the relevant endpoint
pairs have distance $2k$, while every non-shortest path between such endpoints
has length at least $2k+2$. 
The same critical-path argument also gives the corresponding lower bound for approximate distance hopsets in unweighted undirected graphs. We stated the result in reachability terminology because shortcut sets are usually studied in that setting.

We now state the corresponding corollaries for the weighted version of the graph construction by~\cite{kosowski2019hardness} (see \cref{sec:KUV}).
Correctness follows by the same arguments as those for \cref{cor:2-shortcutSetLB} and \cref{cor:unweighted-reachability-hubLabeling}, using the
$\Theta(n^2)$ critical paths in the construction.

\begin{corollary}\label[corollary]{cor:weighted-hopset-LB}
For every constant integer $k\ge 2$, there are infinitely many undirected
weighted layered graphs $G$ with $n$ vertices, $m=O(n^{1+1/k})$ edges, and
$2k+1$ layers, such that any exact hopset with hopbound $2$ for $G$ requires
size $\Omega(n^2)$.
\end{corollary}

\begin{corollary}\label[corollary]{cor:weighted-hubLabeling}
For every constant integer $k\ge 2$, there are infinitely many undirected
weighted layered graphs $G$ with $n$ vertices, $m=O(n^{1+1/k})$ edges, and
$2k+1$ layers, such that any exact distance $2$-hop labeling scheme for $G$
requires total size $\Omega(n^2)$.
\end{corollary}

\subsection{A Sparsity Tradeoff via Endpoint Sparsification}
\label{sec:tradeoff}

We now derive a total-space versus sparsity tradeoff for hub-labeling lower
bounds in graphs with $m=O(n^{1+\alpha})$ edges, for $0\le\alpha<1/k$. The following theorem is a
black-box sparsification step, generalizing a technique of
Cohen et al.~\cite{cohen2003reachability} for reachability labeling lower bounds.
For clarity, we state the theorem for distances in unweighted graphs, but a similar proof applies to  hop-distance in weighted graphs.

We use the following setup. Let $\mathcal N$ be an infinite set of integers.
For every $N\in\mathcal N$, let $G_N$ be an unweighted undirected layered
graph with $N$ vertices, $O(N^{1+\beta})$ edges, and layers
$V_0^N,\ldots,V_{2k}^N$, such that
$|V_0^N|+|V_{2k}^N|=\Theta(N)$. Let
$P_N\subseteq V_0^N\times V_{2k}^N$. We say that this family has endpoint
lower bound $f$ if, for every $N\in\mathcal N$, any labeling scheme that
distinguishes, for every $(u,v)\in P_N$, between
$\dist(u,v)\le 2k$ and $\dist(u,v)\ge 2k+2$ has total label size at least
$f(N)$ on the vertices of $V_0^N\cup V_{2k}^N$.

\begin{theorem}
\label{thm:star-sparsification-labeling}
Let $k\ge 1$ be an integer and let $0\le \alpha<\beta$ be constants. Suppose
there is a family $\{(G_N,P_N)\}_{N\in\mathcal N}$  with endpoint
lower bound $f$\footnote{We assume that $f(cN)=\Theta(f(N))$ for every fixed
$c>0$. All lower-bound functions used below satisfy this property. We thus assume noninteger arguments of $f$ are rounded
down, which does not affect asymptotic complexities.}. Then, for infinitely many integers $n$, there is an
unweighted undirected layered graph $G_n$ with $n$ vertices, $O(n^{1+\alpha})$
edges, and $2k+3$ layers, such that any labeling scheme that distinguishes,
for every vertex pair $u,v$ in $G_n$, between $\dist(u,v)\le 2k+2$ and
$\dist(u,v)\ge 2k+4$ has total label size at least
\[
\Omega\!\left(
n^{\frac{\beta-\alpha}{1+\beta}}\,
f\!\left(n^{\frac{1+\alpha}{1+\beta}}\right)
\right).
\]
\end{theorem}

\begin{proof}
Fix $N\in\mathcal N$ and consider $G_N$, with layers
$V_0,\ldots,V_{2k}$ and  let $P=P_N$. Set
$\gamma=(\beta-\alpha)/(1+\alpha)$ and $t=\lfloor N^\gamma\rfloor$.

$G_n$ is constructed  from $G_N$ by attaching $t$ new leaves to every vertex of
$V_0\cup V_{2k}$. 
For $x\in V_0 \cup V_{2k}$, denote these leaves by
$x^1,\ldots,x^t$.
The vertices inherited from $G_N$ are the inner vertices of $G_n$, and the new leaves are
the outer vertices of $G_n$.

Since $|V_0|+|V_{2k}|=\Theta(N)$, the number of vertices in $G_n$ is
$n=\Theta(Nt)=\Theta(N^{1+\gamma})
=\Theta\!\left(N^{\frac{1+\beta}{1+\alpha}}\right).$
The number of added edges is also $\Theta(Nt)=\Theta(N^{1+\gamma})$. Since
$\gamma\le\beta$, the total number of edges is
\[
m=O(N^{1+\beta})
=
O\!\left(
\left(N^{\frac{1+\beta}{1+\alpha}}\right)^{1+\alpha}
\right)
=
O(n^{1+\alpha}).
\]

Fix $i\in[t]$. For each $(u,v)\in P$, consider the outer pair
$(u^i,v^i)$ where $u\in V_0$ and $v\in V_{2k}$. Every path from $u^i$ to $v^i$ must pass through $u$ and $v$, and
therefore
\[
\dist_{G_n}(u^i,v^i)=\dist_{G_N}(u,v)+2.
\]
Thus distinguishing $\dist_{G_N}(u,v)\le 2k$ from
$\dist_{G_N}(u,v)\ge 2k+2$ is equivalent, on these outer pairs, to
distinguishing $\dist_{G_n}(u^i,v^i)\le 2k+2$ from
$\dist_{G_n}(u^i,v^i)\ge 2k+4$.

Suppose that a labeling scheme for ${G_n}$ distinguishes between these two cases for every
vertex pair. Let
$
X_i=\{u^i\mid u\in V_0\}\cup \{v^i\mid v\in V_{2k}\}.
$
Using only the labels of vertices in $X_i$, we obtain a labeling scheme for the vertices of $V_0\cup V_{2k}$ in $G_N$ by assigning to each $u\in V_0$ the label
of $u^i$, and to each $v\in V_{2k}$ the label of $v^i$. 
The decoder simulates the decoder for ${G_n}$ on the pair $(u^i,v^i)$. Since
$\dist_{G_n}(u^i,v^i)=\dist_{G_N}(u,v)+2$, it outputs
$\dist_{G_N}(u,v)\le 2k$ exactly when the simulated decoder outputs
$\dist_{G_n}(u^i,v^i)\le 2k+2$, and otherwise outputs
$\dist_{G_N}(u,v)\ge 2k+2$.
Thus, the total size of the labels on $X_i$ is at
least $f(N)$.

The sets $X_i$ are pairwise disjoint over $i\in[t]$. Therefore, the total label
size in ${G_n}$ is at least
\[
\Omega(t\cdot f(N))
=
\Omega\!\left(
n^{\frac{\beta-\alpha}{1+\beta}}\,
f\!\left(n^{\frac{1+\alpha}{1+\beta}}\right)
\right),
\]
where we used
$t=\Theta(N^\gamma)=\Theta(n^{(\beta-\alpha)/(1+\beta)})$ and
$N=\Theta(n^{(1+\alpha)/(1+\beta)})$.
\end{proof}

\cref{thm:star-sparsification-labeling} is stated for general labeling
schemes to emphasize that the sparsification step is not specific to hub
labelings. A similar argument also applies to hub-labeling schemes, where this time $f$ represents the total hub sizes instead of total label sizes. The idea is similar: 
for each $i\in [t]$, the same hubs stored for the vertices of $X_i$ can be used to construct a valid hub labeling scheme for the vertices of $V_0\cup V_{2k}$ in $G_N$ by transferring the hubs of a leaf $u^i$ to its center $u$, and similarly the
hubs of $v^i$ to $v$, while changing the stored hub distances appropriately.
Thus, as in the proof for general labeling schemes, the total hub
sizes in ${G_n}$ expands by a factor of $t$.

Applying the hub-labeling version of \cref{thm:star-sparsification-labeling} to undirected
approximate distance version in
\cref{cor:unweighted-reachability-hubLabeling}, with $\beta=1/k$ and $f(N)=\Omega\!\left(N\cdot r_3(N^{1/k})^k\right)$, gives the
following tradeoff.

\begin{corollary}
\label[corollary]{cor:approx-hubLabeling-tradeoff}
Fix a constant integer\footnote{We state the result for $k\ge 2$ since the $(2k+1)$-layered graphs we were interested in until now only gave meaningful lower bounds for $k\ge 2$. However, this result also similarly extends to $k=1$. For the case of $k=1$ and $\alpha=0$, a stronger simple $4$-layered graph construction is known to give a $\Omega(n^{3/2})$ lower bound~\cite{BermanRR10}. Since the graph is $4$-layered ($3$ edge layers), it even gives a stronger stretch guarantee of any stretch value strictly smaller than $5/3$.} $k\ge 2$ and a constant $0\le \alpha<1/k$. There are
infinitely many unweighted undirected graphs $G$ with $n$ vertices and
$m=O(n^{1+\alpha})$ edges such that any hub labeling scheme with stretch
strictly smaller than $1+1/(k+1)$ requires total size at least
\[
\Omega\!\left(
n^{\frac{\beta-\alpha}{1+\beta}}\,
f\!\left(n^{\frac{1+\alpha}{1+\beta}}\right)
\right)
=
\Omega\!\left(
n^{\frac{1/k-\alpha}{1+1/k}}\,
n^{\frac{1+\alpha}{1+1/k}}\,
r_3\!\left(n^{\frac{1+\alpha}{k+1}}\right)^k
\right)
=
\Omega\!\left(
n\cdot r_3\!\left(n^{\frac{1+\alpha}{k+1}}\right)^k
\right)
=
\Omega\!\left(
\frac{
n^{1+\frac{k(1+\alpha)}{k+1}}
}{
2^{O_k(\sqrt{\log n})}
}
\right)
=
n^{2-\frac{1-k\alpha}{k+1}-o(1)}.
\]
In particular, for $\alpha=0$, this gives graphs with $m=O(n)$ for which any
such hub labeling scheme requires total size 
\[
n^{2-\frac{1}{k+1}-o(1)}.
\]
\end{corollary}

By \cref{lem:two-hop-labels-hopsets-equivalence}, the same bounds imply
corresponding lower bounds for approximate distance hopsets with hopbound $2$.
The weighted exact-distance hub-labeling version of
\cref{thm:star-sparsification-labeling}, applied to
\cref{cor:weighted-hubLabeling}, gives the corresponding bounds for
weighted exact-distance hopsets and weighted hub-labeling schemes.
By Corollary~\ref{cor:weighted-hubLabeling}, in the weighted case we have endpoint lower bound $f(N)=\Omega(N^2)$, 
and so the corresponding Corollary~\ref{cor:approx-hubLabeling-tradeoff} for weighted graphs would give a lower bound of
$$\Omega\!\left(
n^{\frac{\beta-\alpha}{1+\beta}}\,
f\!\left(n^{\frac{1+\alpha}{1+\beta}}\right)
\right) 
= \Omega\!\left(
n^{\frac{1/k-\alpha}{1+1/k}}\,
n^{\frac{2(1+\alpha)}{1+1/k}}\right)
= \Omega\!\left(n^{2-\frac{1-k\alpha}{k+1}}\right).
$$

\subsection{A Matching Upper Bound for 2-Hop Shortcut Sets}\label{sec:upperBounds}
We complement the results in \cref{sec:consequences} and \cref{sec:tradeoff} with the following upper bound (see~\cite{BermanRR10, goldstein2017conditional}).
We phrase the theorem in the terminology of shortcut sets, but the same proof (with very minor changes) works for directed weighted exact-distance hopsets.
By \cref{lem:two-hop-labels-hopsets-equivalence}, the following theorem also implies hub labeling scheme constructions.

\begin{theorem}\label{thm:upper}
    For constant integer $k\geq1$, 
    given a directed graph $G$ with $n$ vertices, $m = O(n^{1+1/k})$ edges, and diameter of at most $2k$, then $G$ has a shortcut set $H$ of size $|H| = O(m n^{1-1/k})$ such that the diameter of $G \cup H$ is at most $2$.
\end{theorem}

\begin{proof}
    The construction of the shortcut set $H$ is based on a heavy-light technique\footnote{A similar implementation of the same heavy-light technique was suggested in~\cite{goldstein2017conditional} for a construction for a distance oracle variant of the $k$-Reachability problem, and in~\cite{BermanRR10} for constructing transitive closure spanners, an object strongly related to shortcut sets.}. 
    For a vertex $v$ in $G$, denote the (in and out) neighbors of $v$ in $G$ with $N(v, G)$, let $R^{-}(v, G)$ be the vertices that reach $v$ in $G$, and let $R^{+}(v, G)$ be the vertices that $v$ reaches in $G$. 
    We define $v$ to be a heavy vertex if $|N(v, G)| \geq n^{1/k}$, and define $v$ as a light vertex otherwise. 
    For every heavy vertex $v$, add to the shortcut set $H$ a shortcut from each vertex in $R^{-}(v, G)$ to $v$, and a shortcut from $v$ to each vertex in $R^{+}(v, G)$. Since there are at most $O(m/n^{1/k})$ heavy vertices, the number of shortcuts added is at most $O(nm/n^{1/k}) = O(mn^{1-1/k})$. 
    
    Consider the graph $G' \subseteq G$, which is defined to be the subgraph of $G$ induced by the light vertices in $G$. For each edge $e=(u,v) \in G'$, denote the vertices in $G'$ that reach $u$ using at most $k-1$ edges with $I(e) = \{x \in R^{-}(u, G')\mid d_{G'}(x,u) \leq k-1\}$, and the vertices in $G'$ that $v$ reaches using at most $k-1$ edges as $O(e) = \{x \in R^{+}(v, G')\mid d_{G'}(v,x) \leq k-1\}$. 
    For each $e=(u,v) \in G'$, add to $H$ a shortcut from all vertices in $I(e)$ to $v$, and add a shortcut from $u$ to all the vertices in $O(e)$. 
    Notice that now for any two vertices $a,b$ such that $\dist_{G'}(a,b) \leq k$, it holds that $(a,b) \in H$. 
    For each edge in $G'$, the number of shortcuts added to $H$ is $O((n^{1/k})^{k-1}) = O(n^{(k-1)/k})$. 
    Since the number of edges in $G'$ is at most $m$, the total edges added to $H$ is $O(mn^{1-1/k})$, which is the same bound as in the first step of the algorithm, so $|H|=O(mn^{1-1/k})$.

    For correctness, consider a pair of vertices $x$ and $y$ in $G$ such that $d_G(x,y)\le2k$. 
    If there exists a shortest path in $G$ from $x$ to $y$ that passes through a heavy vertex $h$, then there exist shortcuts $(x,h)$ and $(h,y)$ in $H$, which makes $\dist_{G\cup H}(x,y) \le 2$. 
    Otherwise, all vertices on every shortest path between $x$ and $y$ are light.
    Consider such a shortest path $P$ from $x$ to $y$ in $G'$. 
    Since $\dist_{G'}(x,y) \leq 2k$, there must be a vertex $\ell \in P$ such that $\dist_{G'}(x,\ell) \leq k$ and $\dist_{G'}(\ell,y) \leq k$. 
    The shortcuts added in the second phase of the construction guarantee that the shortcuts $(x,\ell)$ and $(\ell,y)$ exist in $H$, and thus, $d_{G\cup H}(x,y) \leq 2$. 
    
    We conclude that by adding the shortcut set $H$ of size $|H| = O(mn^{1-1/k})$ to $G$, the diameter of $G \cup H$ is at most $2$.
\end{proof}

\subsection{Discussion - Comparison of the Upper and Lower Bounds}\label{sec:discussion}
The upper bound from \cref{thm:upper} complements our lower bounds from this section in a tight way in the following sense. 
For a constant $k\ge 2$, consider an $n$-vertex directed weighted layered graph with $2k+1$ layers and $\Theta(n)$ vertices in each layer.
For graphs with $m=\Theta(n^{1+1/k})$, the lower bound from \cref{cor:weighted-hubLabeling} implies that no non-trivial hub labeling scheme with total space $o(n^2)$ exists for such graphs. 
On the other hand, when $m=o(n^{1+1/k})$ the (weighted graph, hub-labeling version of) \cref{thm:upper} implies that a non-trivial  hub labeling scheme (with total space $o(n^2)$) exists for such graphs. 
Similarly, for the density regime $m=O(n)$ (and $k\ge 3$), we obtain an upper bound of $O(n^{2-1/k})$ by \cref{thm:upper}, and a lower bound of $\Omega(n^{2-1/k})$ by the result which appears in the discussion immediately following \cref{cor:approx-hubLabeling-tradeoff}\footnote{Notice that \cref{cor:approx-hubLabeling-tradeoff} and the corresponding result for weighted graphs concern graphs with $2k+3$ layers and stretch strictly smaller than $1+1/(k+1)$, as opposed to graphs with $2k+1$ layers, and so for a lower bound for $2k'+1$-layered graphs we use $k'=k+1$.}.

For unweighted graphs, however, a sub-polynomial gap remains: for graphs with $m=O(n^{1+1/k})$  we  show a lower bound of
$
\Omega\!\left(n\cdot r_3(n^{1/k})^k\right)
\ge n^2/{2^{\Theta_k(\sqrt{\log n})}},
$ (\cref{cor:unweighted-reachability-hubLabeling}),
but we are unaware of any $o(n^2)$ upper bound.
We note that our lower bound construction for the $m=O(n^{1+1/k})$ regime uses maximum degree $O(r_3(n^{1/k}))=o(n^{1/k})$, and so the upper bound construction from \cref{thm:upper} for this graph automatically gives an upper bound of $o(n^2)$. In other words, to obtain a quadratic lower bound for the $m=O(n^{1+1/k})$ setting, one would have to utilize maximum degree $\Omega(n^{1/k})$.

Thus, it remains an interesting open problem to determine the exact asymptotic bounds for shortcut sets which reduce the diameter to $2$ in graphs with $m=O(n^{1+1/k})$ and diameter $2k$.
A similar open problem remains for the case $m=O(n)$ (and $k\ge 3$), where we also obtain near tight bounds up to subpolynomial factors\footnote{For unweighted undirected graphs with $m=O(n)$, Kosowski, Uzna\'nski, and Viennot~\cite{kosowski2019hardness} show a 2-hop hub labeling upper bound of $O\left(\frac{n^2}{\mathrm{RS}(n)^{1/7}}\right)=o(n^2)$, where $\mathrm{RS}(n)$ is the Ruzsa--Szemer\'edi function, known to satisfy $2^{\Omega(\log^* n)}
\leq \mathrm{RS}(n)
\leq 2^{O(\sqrt{\log n})}$. 
It is possible that similar techniques could be used to improve the upper bound in our shortcut set setting where $G$ satisfies $m=O(n)$ and has diameter $2k$ for constant $k\ge 3$, from $O(n^{{2-1/k}})$ to $o(n^{2-1/k})$.}. 
For $m=O(n^{1+\alpha})$ and $k\ge3$, where
$0<\alpha<1/k$, a polynomial gap remains: we achieve an upper bound of $O(n^{2-1/k+\alpha})$ and a lower bound of $\Omega\!\left(
n^{2-\frac{1-(k-1)\alpha}{k}-o(1)}
\right)
=
n^{2-1/k+\frac{k-1}{k}\alpha-o(1)}$, 
and so the polynomial gap is $n^{\alpha/k+o(1)}$.


\paragraph{Acknowledgments.}

The authors used OpenAI's ChatGPT, primarily models in the GPT-5 family, for language editing, and GPT-6.1 for assistance with preparing the TikZ code for the figure illustrating the graph construction. The authors take full responsibility for all content.

\appendix

\section{Proof of \cref{thm:sumindex-constant-factor}}
\label{appndx:missingProofs}

\begin{proof}[Proof of \cref{thm:sumindex-constant-factor}]

Let $\Pi$ be a protocol for Sum-Index on strings of length $n$, with communication complexity $\SI(n)$.
We construct a protocol for Sum-Index on strings of length $kn$.

Let $S\in\{0,1\}^{kn}$ be the common input string. We partition $S$ into $k$ consecutive blocks of length $n$. For every $t\in\lrb k$, define $S^{(t)}\in\{0,1\}^n$ by 
\[
S^{(t)}[x]=S[tn+x]
\qquad\text{for every }x\in\lrb n.
\]

Alice receives $i\in\lrb{kn}$ and Bob receives $j\in\lrb{kn}$. Write $i=an+i_0$ and $j=bn+j_0$, where $a,b\in\lrb k$ and $i_0,j_0\in\lrb n$.
For every $t\in\lrb k$, Alice and Bob simulate the protocol $\Pi$ on the common string $S^{(t)}$, with Alice's index $i_0$ and Bob's index $j_0$, and send all the resulting messages to the referee. In addition, Alice sends $a$ and $i_0$, and Bob sends $b$ and $j_0$.

The referee defines $x=(i_0+j_0)\bmod n$ and a carry bit $c$ 
as,
\[
c=
\begin{cases}
0, & \text{if } i_0+j_0<n,\\
1, & \text{if } i_0+j_0\ge n,
\end{cases}
\]
so $i_0+j_0=cn+x$. 
In addition, the referee sets $t^*=(a+b+c)\bmod k$. The referee returns $S^{(t^*)}[x]$, which we claim is the correct bit $S[(i+j)\bmod kn]$.
Indeed,
\[
i+j
= (an+i_0)+(bn+j_0)
= (a+b)n+(i_0+j_0)
= (a+b)n+(cn+x)
= (a+b+c)n+x.
\]
Thus,
\[
S[(i+j)\bmod kn]
=S[((a+b+c)n+x)\bmod kn]
=S[(((a+b+c)\bmod k)n+x)]
=S[t^*n+x]
=S^{(t^*)}[x].
\]
The communication complexity is $k\cdot \SI(n)+ O(\log n) + O(\log k) = k\cdot \SI(n)+O(\log n)$. By \cref{thm:sumindex-lower-bound}, $\SI(n)=\Omega(\sqrt n)$, so the term $O(\log n)$ is dominated by $\SI(n)$. Thus, $\SI(kn)=O(k\cdot \SI(n))$.
\end{proof}

\section{An implicit result in~\cite{kosowski2019hardness}}\label{sec:KUV}

In this section we explain an implicit result of a  weighted construction underlying the lower bound of Kosowski, Uzna\'nski, and Viennot~\cite{kosowski2019hardness}. 
We state the construction in a form that parallels \cref{thm:avgfree-layered-CP}, since this makes the comparison with our unweighted construction transparent. 
We also record the corresponding labeling lower bound, which follows by the same Sum-Index reduction template used in the main body.

\begin{lemma}[\cite{kosowski2019hardness}]\label[lemma]{thm:weighted-layered-CP}
For every constant integer $k\ge 1$ and even integer $N\ge 2$, there exists an undirected weighted layered graph $H$ with the following properties.

\begin{enumerate}
    \item The graph has $2k+1$ layers where each layer consists of $N^k$ vertices where each vertex corresponds to a unique vector in $\lbb{N}\rbb^k$.
    We denote a vertex in $H$ as $(\ell, \vec v)$, where $\ell\in \lrb {2k+1}$ is the layer index and $\vec v\in \lbb{N}\rbb^k$ is the vector.
    
    Thus, the total number of vertices is
        $n=(2k+1)N^k.$

    \item 
    The total number of edges is $m = 2k N^{k+1} =  O(n^{1+1/k})$ edges.


    \item For every pair of vectors $\vec x,\vec z\in (2\lbb{N/2}\rbb)^k$, there is a unique shortest path, denoted as $P_{\vec x, \vec z}$ between the vertex $(0,\vec x)$ 
    and the vertex $(2k,\vec z)$. 
    Moreover, $P_{\vec x, \vec z}$ uses exactly $2k$ edges and passes through the unique vertex
    $(k,(\vec x+\vec z)/{2})$
    in layer $k$.
    
    We denote the set of such unique shortest paths as $\mathcal{P} =\{P_{\vec x, \vec z} \mid \vec x,\vec z\in (2\lbb{N/2}\rbb)^k\}$. We have
\[
    |\mathcal{P}|
    =
    \left(\frac{N}{2}\right)^{2k}
    =
    \frac{n^2}{2^{2k}(2k+1)^2}
    =
    \Theta(n^2).
\]
\end{enumerate}
\end{lemma}

Using exactly the same proof idea in Kosowski, Uznanski and Viennot~\cite{kosowski2019hardness}, the following theorem follows.

\begin{theorem}[\cite{kosowski2019hardness}]\label{thm:weighted-hop-distance-labeling-sumindex} For every constant integer $k\ge 1$, there are infinitely many integers $n\ge1$ such that every labeling scheme that returns exact $2k$-hop distances on undirected weighted graphs with $n$ vertices has maximum label length $\Omega(\SI(n))$ on some graph $G$ with $m=O(n^{1+1/k})$ edges. \end{theorem}

\begin{proof}

Let $s\ge 1$ be an integer, and consider the Sum-Index problem on a common string of length $s^k$. 
We design a protocol for the Sum-Index problem using a distance labeling scheme for a graph from the graph family we now describe.

Let $G_0$ be the weighted graph from \cref{thm:weighted-layered-CP} with parameters $k$ and $N=2s$. Thus, $G_0$ has $n=(2k+1)(2s)^k$ vertices, $m=O(n^{1+1/k})$ edges, $2k+1$ layers, and each layer consists of vertices corresponding to the vectors in $\lbb{2s}\rbb^k$.

For every string $X\in\{0,1\}^{s^k}$, define a graph $G_X$ on the same vertex set as $G_0$ as follows. For every $\vec y\in\lbb{2s}\rbb^k$ such that $X[\repr_s(\vec y)\bmod s^k]=0$, delete all edges incident to $(k,\vec y)$. 
The rest of the edges of $G_0$ remain unchanged.
The idea is that the middle-layer vertices of $G_X$ encode the string $X$ in a way that the referee will use the labels of the endpoints of a critical path in $G_X$ to recover the specific bit the referee needs in $X$.

We now describe the protocol for the Sum-Index problem. 
To define the protocol, 
fix any labeling scheme that returns exact $2k$-hop distances for all graphs in $\{G_X \mid X\in\{0,1\}^{s^k}\}$. Let $L$ be the maximum label length assigned by this scheme to any vertex of any graph in the family. We will show a lower bound for $L$.

On input, Alice and Bob receive a common string $S\in\{0,1\}^{s^k}$. Additionally, Alice receives an index $a\in\lrb{s^k}$, Bob receives an index $b\in\lrb{s^k}$, and the referee must output $S[(a+b)\bmod s^k]$.

Let $\vec i,\vec j\in\lbb{s}\rbb^k$ be the base-$s$ representations of $a$ and $b$, respectively, so that $\repr_s(\vec i)=a$ and $\repr_s(\vec j)=b$, and consider the vertices $u=(0,2\vec i)$ and $v=(2k,2\vec j)$.

Since $\vec i,\vec j\in\lbb{(2s)/2}\rbb^k$, we have $2\vec i,2\vec j\in(2\lbb{(2s)/2}\rbb)^k$. Therefore, by \cref{thm:weighted-layered-CP}, there is a unique shortest path $P_{2\vec i,2\vec j}$ between $u$ and $v$ in $G_0$. The path $P_{2\vec i,2\vec j}$ has exactly $2k$ edges and passes through $w=(k,(2\vec i+2\vec j)/2)=(k,\vec i+\vec j)$.
Moreover, we have
\[\repr_s(\vec i+\vec j)\bmod s^k=(\repr_s(\vec i)+\repr_s(\vec j))\bmod s^k=(a+b)\bmod {s^k}.\]
Consider the graph $G_S$. If $S[(a+b)\bmod s^k]=1$, then the edges incident to $w$ remain in $G_S$. Thus, the path $P_{2\vec i,2\vec j}$ also remains in $G_S$, and therefore, $d_{G_S}^{(2k)}(u,v)=d_{G_0}^{(2k)}(u,v)$. 
Otherwise, all edges incident to $w$ are deleted, and so the unique shortest path $P_{2\vec i,2\vec j}$ is not included in $G_S$, and $d_{G_S}^{(2k)}(u,v)>d_{G_0}^{(2k)}(u,v)$.

Alice sends the label of $u$ in $G_S$ together with $a$, and Bob sends the label of $v$ in $G_S$ together with $b$. The referee recovers $d_{G_S}^{(2k)}(u,v)$, and computes $d_{G_0}^{(2k)}(u,v)$ from $a$, $b$ and the fixed graph $G_0$.

Finally, the referee returns
\[
S[(a+b)\bmod s^k]=
\begin{cases}
1, & \text{if } d_{G_S}^{(2k)}(u,v)=d_{G_0}^{(2k)}(u,v),\\[2mm]
0, & \text{if } d_{G_S}^{(2k)}(u,v)>d_{G_0}^{(2k)}(u,v).
\end{cases}
\]

By the analysis above, we have a correct protocol for Sum-Index on strings of length $s^k$.
The communication complexity of both Alice and Bob is at most $L+O(k\log s)$, and therefore,
\[
\SI(s^k)\le L+O(k\log s).
\]

Since $k$ is a constant, by \cref{thm:sumindex-lower-bound}, the term $O(k\log s)$ is insignificant, and $n=(2k+1)(2s)^k=\Theta((2s)^k)=\Theta(s^k)$. It follows by \cref{thm:sumindex-constant-factor} that 
\[
L=\Omega\!\left(\SI(s^k)\right)
=\Omega\!\left(\SI((2s)^k)\right)
=\Omega\!\left(\SI(n)\right).
\]
Thus, every labeling scheme that returns exact $2k$-hop distances assigns some vertex a label of length $\Omega(\SI(n))$ in some graph $G_X$ in the graph family, which completes the proof.
\end{proof}


\bibliography{refs}

\end{document}